\documentclass[12pt]{article}

\usepackage{graphicx}
\usepackage{amssymb}
\usepackage{epstopdf}
\usepackage{amsfonts}
\usepackage{amsmath}
\usepackage{bm}
\usepackage{mathrsfs}
\usepackage{subfigure}

\usepackage{tabularx}
\usepackage{xcolor}
\usepackage{authblk}

\def\be{\begin{equation}}
\def\te{\end{equation}}
\def\ee{\end{equation}}

\def\bea{\begin{eqnarray}}
\def\tea{\end{eqnarray}}
\def\eea{\end{eqnarray}}

\def\be{\begin{equation}} \def\ee{\end{equation}} \def\te{\end{equation}} \def\bea{\begin{eqnarray}}


\numberwithin{equation}{section}

\begin{document} 

\title{\Large  Quantum Capacity and Vacuum Compressibility of Spacetime: Thermal Fields}

% Paper II: Quantum Capacity and Vacuum Compressibility of the Universe: Nonequilibrium Fields 

%\author{Hing-Tong Cho, Jen-Tsung Hsiang and Bei-Lok Hu}
\author[1]{Hing-Tong Cho\thanks{htcho@mail.tku.edu.tw }}
\author[2]{Jen-Tsung Hsiang\thanks{cosmology@gmail.com}} 
\author[3]{Bei-Lok Hu\thanks{blhu@umd.edu}}
\affil[1]{Department of Physics, Tamkang University, Tamsui, New Taipei City 251301, Taiwan, ROC}
\affil[2]{Center for High Energy and High Field Physics, National Central University, Taoyuan 320317, Taiwan, ROC}
\affil[3]{Maryland Center for Fundamental Physics and Joint Quantum Institute,\\ University of Maryland, College Park, Maryland 20742-4111 U.S.A.}

\date{\small (\today)}
 
\maketitle

\begin{abstract}  
An important yet  perplexing result from work in the 90s and 00s is  the near-unity value of the ratio of fluctuations in the vacuum energy density of quantum fields to the mean in a collection of generic spacetimes. 
%It is a known fact that the magnitude of energy density fluctuations is comparable to the mean in many quantum systems, from thermal fields in Minkowski and Casimir geometry to the Einstein and  de Sitter universes and more.  
This was done by way of  calculating the noise kernels which are the correlators of the stress-energy tensor of quantum fields.  In this paper we revisit this issue via a quantum thermodynamics approach, by calculating two quintessential  thermodynamic quantities: the heat capacity and the quantum compressibility of some model geometries filled with a quantum field at  high and low temperatures.  This is because heat capacity at constant volume  gives  a measure of the fluctuations of the energy density to the mean. When  this ratio  approaches or exceeds unity, the validity of the canonical distribution is called into question.  Likewise, a system's compressibility at constant pressure is a criterion for the validity of grand canonical ensemble.  We derive the free energy density and, from it, obtain the expressions for these two thermodynamic quantities for  thermal and quantum fields in 2d Casimir space, 2d Einstein cylinder and  4d ($S^1 \times S^3$ ) Einstein universe. To examine the dependence on the dimensionality of space, for completeness, we have also derived these thermodynamic quantities for the Einstein universes with even-spatial dimensions: $S^1 \times S^2$  and $S^1 \times S^4$.  With this array of spacetimes we can investigate the thermodynamic stability of quantum matter fields in them  and make some qualitative observations on the  compatibility condition for the co-existence between quantum fields and spacetimes, a fundamental issue in the  quantum and gravitation conundrum.  
\end{abstract}
\newpage
\tableofcontents
\newpage

%Use thermal field in Minkowski geometry (hot flat space) to establish the correspondence between noise kernel (energy and momentum density fluctuations) and heat capacity and compressibility. First via quantum thermodynamics, then via stress tensor correlators.   Then do the thermodynamics of quantum field fluctuations for a Casimir (// plate) geometry. The compressibility of thermal fields  will give us some idea of what roles quantum field and geometry play  respectively in contributing to the thermodynamic instability condition.

\section{Introduction}

Three major elements are embedded in the theme explored in this paper:  quantum fields,  spacetime, and thermodynamics.  The  first two describe how quantum matter is affected by  spacetime, its geometry and topology, and how it steers the  dynamics of spacetime,  although the latter issue involving backreaction is not explored here.   The relation between matter and spacetime dynamics is of course underwritten by Einstein's general relativity theory. With quantum matter as source, one needs a theory of quantum fields in curved spacetime \cite{BirDav,ParTom},  which is a test field limit of semiclassical gravity theory  \cite{HuVer}, where the dynamics of the matter field and the spacetime are treated self-consistently.   The first and third elements refer to the thermodynamics of quantum fields, a familiar subject,  by way of thermal field theory.  In this paper we shall only consider quantum fields under equilibrium conditions, which can exist for static spacetimes, and be treated by finite temperature quantum field theory.  Later, when we tackle cosmological issues, we shall  call upon the more  challenging subject of nonequilibrium quantum field theory \cite{CalHu08} and nonequilibrium quantum thermodynamics \cite{HHNEqFE}.  The second and third elements bear on the thermodynamics of spacetime, which is also an old topic, ranging from the thermodynamic properties of classical matter, such as  gravitating systems having  negative heat capacity \cite{LynBell,Thirring}  to  black hole mechanics \cite{BCH} and thermodynamics \cite{Bek} and,  when quantum physics is taken into the consideration,  the famous Bekenstein-Hawking  entropy \cite{Haw} and its deep physical meaning.  

All told, these are the issues one needs to consider behind the grander views of  spacetime thermodynamics \cite{Jacobson},  general relativity as geometro-hydrodynamics \cite{GRhydro},  and emergent gravity \cite{Volovik,E/QG,Sindoni}, where the large scale structure and dynamics of spacetime can be phrased in  thermodynamic  or hydrodynamic terms.   
 What we want to accomplish in  this and a sequel paper \cite{CHH2} is more restricted in scope, aiming at firing some solid bricks toward building  this superstructure.

\subsection{Energy and Pressure Densities of Quantum Fields and Fluctuations}  

One fundamental aspect of special interest to us is fluctuation phenomena in quantum matter fields and how they influence the dynamics and the thermodynamics of spacetimes.   We describe their importance and plan for two ways to approach them, by calculating two important thermodynamic quantities, the heat capacity and the vacuum compressibility here and later, via noise kernels in quantum field theory in curved spacetime. 

\paragraph{Importance of Quantum Field Fluctuations Phenomena}

 Quantum field fluctuations phenomena have both fundamental theoretical and practical application values. The simplest example of a vacuum polarization effect is perhaps the Casimir effect \cite{Casimir,Ford,DowCri}:  e.g., an attractive force between two conducting plates, or a repulsive force in a  conducting sphere. When the plates are moved rapidly, particles are created from the vacuum. Dynamical Casimir effect \cite{Dod} stems from the amplification of vacuum fluctuations by changing boundaries \cite{Moore,FulDav,DavFul}. Cosmological particle creation \cite{Par69,Zel70} shares the same physics, the driving agent being the expanding universe. The theoretical basis is quantum fields in curved spacetime \cite{BirDav,ParTom}.
Quantum effects in the early universe  \cite{ZelSta71,HuPar77,HuPar78,HarHu79,FHH79,Star80,Guth,CalHu87,CalHu94,HuMatELE,HuSinFDR,CamVer94,CamVer96} invoking  semiclassical  and stochastic gravity theories \cite{HuVer} are believed to have played a decisive role in shaping our present universe. 

The purpose of our present investigation is to knit a thermodynamic picture of these quantum field fluctuation phenomena.  Our plan is to take a two-prong approach to explore these issues, one pursued in this and a sequel paper  uses thermal field theory and quantum field theory  in curved spacetimes, via free energy density and partition functions; the other approach  invokes the noise kernel in stochastic gravity \cite{HuVer}. The noise kernel is the vacuum expectation value of the stress-energy bitensor \cite{PH00} or the stress-tensor two point functions (correlators).    The 00 component and the ii components give respectively the fluctuations in the energy and momentum density.  An interesting result from works in the 90s and 00s is  the near-unity value of the ratio of fluctuations in the vacuum energy density to the mean. This quantity $\Delta$  has been calculated for   thermal fields in Minkowski and Casimir geometries,  the Einstein universe, de Sitter and anti-deSitter spacetimes and more \cite{KuoFor,PH97,PH00,PH03,ChoHu11,ChoHu12,ChoHu15}.   
This is a plain yet intriguing result. It may  hold some deeper meanings in how quantum fields (matter contents) co-exist with spacetime (geometry and topology). Possible implications on the condition of our universe is a theme worth further explorations \cite{HuCritU}.

In this paper we wish to shed some light on this issue by way of  a thermodynamic approach,  by computing  two essential thermodynamic quantities, the heat capacity and the adiabatic compressibility.  

\paragraph{Thermodynamics of quantum fields reflecting the properties of spacetime }

One commonly studied thermodynamic quantity is the heat capacity, well-known as a measure of the magnitude of energy fluctuations. When it diverges  it spells the breakdown of canonical ensembles in thermodynamics.

In general the heat capacity is negative \cite{LynBell,Thirring} for gravitating systems (see, e.g., \cite{Mario} and references therein). That is why when one applies thermodynamics in a canonical ensemble setting one needs to add the condition that the system, e.g, a black hole,  needs to be placed in a box or in an AdS space. We need to separate two effects simultaneously affecting the heat capacity, one due to quantum fluctuations, the other due to gravity.  
To explore gravitating systems without an artificially imposed boundary we need to use microcanonical ensembles \cite{BroYor} where the starting point would be the number of accessible states to an isolated system at a certain energy. 

The other thermodynamic quantity of equal importance but lesser studied in  field theory is compressibililty,   isothermal at finite temperature  and adiabatic compressibility at zero temperature.  As is evident from its definition, it measures how compressible the quantum field is,  which depends on the curvature and topology of the space they live in.  In fact when we refer to, say, the capacitance of a parallel plate capacitor in electrostatics, we usually just give a  formula in terms of geometric measures, such as the area of the plates and the distance between them. It is implicit that we are talking about electric charges and fields. It is in this sense that we refer to the capacity and compressibility of spacetimes in terms of how a quantum field and its fluctuations behave. Since heat capacity is studied widely we shall focus more on the latter.  The term `quantum capacity' in the title of this paper refers to the heat capacity of quantum fields, including high and low temperatures;  and when we say `vacuum compressibility' we refer to adiabatic compressibility which can be defined without any notion of heat.

\subsection{Physical contexts of Quantum Capacitance and Negative Compressibility}

To better appreciate the meaning and significance of these thermodynamic quantities it is useful to see how quantum capacitance and vacuum compressibility are defined and used in some more familiar physical systems. We give three examples here: nuclear ``liquid drop" model, 2-dim electron gas and graphene nanotubes. Nuclear collective model is a good analog because the Hamiltonian  quadrature has the same mathematical form as the metric of the mixmaster universe \cite{MisMix,Hu73}, a compact anisotropic spacetime,   where compressibility has some intuitive meaning.  2-dim electron gas has similar thermodynamic behavior as the gravitational field, in that they are both systems of long range interactions and with negative heat capacities (see, e.g.,  \cite{Mario} and references therein).  In graphene we can see how quantum capacitance enters due to the dependence on the number density. Besides, such systems have active experiments closely tracking the theoretical analysis. 

\paragraph{Nuclear compressibility in quantum hadrodynamics}

The compressibility of a nucleus is directly related to the surface  energy --  a high compressibility implies a stiff surface which will have a large contribution
to the total energy of the deformed nucleus. For example, Price and  Walker \cite{Price} 
developed a self-consistent, relativistic theory of deformed nuclei based on quantum hadrodynamics and the finite Hartree approximation, and have applied this theory to the calculation of deformed orbitals in various light nuclei. 
  They even went into details in explaining why a relativistic mean field calculation based on quantum hadrodynamics \cite{SerWal}  gives a larger value of  compressibility than non-relativistic calculations.   Inasmuch as the nuclear collective model sharing a similar mathematical structure (Hamiltonian quadratic form) as the mixmaster universe (diagonal Bianchi type IX metric), the Hartree approximation in seeking self-consistent solutions are similar to the structure and procedures in   semiclassical gravity theory. 

\paragraph{What does the negative compressibility of a 2-dim electron gas reveal?}

For two ­dimensional electron systems   at high magnetic fields, interaction effects have spectacular transport con­sequences, e.g., the fractional quantum Hall effect.  The importance of exchange and correlation contributions to the total energy, and hence the thermodynamics, of interacting electron systems has long been theoretically appreciated. For example, in the low-density regime, where interactions dominate the kinetic energy, the ex­change energy alone is sufficient to produce a negative compressibility for the electron gas --  the inverse of the compressibility is given by $\kappa^{-1} = N^2 \partial \mu /\partial n $, where  $\mu$ is the chemical potential and $n$ is the number density.  
Eisenstein et al  \cite{Eisenstein}  reported on a new experimental technique whereby they can directly extract both the sign and the magnitude of $\kappa$ as a function of electron density $n$.  Regions of nega­tive   $\kappa$ are observed at both zero and high magnetic field extreme quantum limit.  Observation of these compressibility features constitutes strong thermodynamic evidence for existence of the dilute quasiparticle gases central to theory of the fractional quantum Hall effect.
This is one of many examples of how (macroscopic) thermodynamic quantities may reveal some important attributes of the underlying (microscopic) constituents and their interactions. 

\paragraph{Quantum capacitance of carbon nanotube contains information about its ground state}  

We can use the words of Ilani et al \cite{Ilani} to illustrate the importance of quantum capacitance: ``The electronic capacitance of a one-dimensional system such as a carbon nanotube is a thermodynamic quantity that contains fundamental information about the ground state.  It is composed of an electrostatic component describing the interactions between electrons and their correlations, and a kinetic term given by the electronic density of states."  The measurements in their experiments suggest the existence of a negative capacitance, which is predicted to exist in one dimension as a result of interactions between electrons.   This lower dimensional system offers a good example to see how quantum effects enter into the thermodynamic quantities. ``The capacitance of a classical conductor is determined solely by its geometry. When charged, the electrons distribute in space in a manner that minimizes their electrostatic energy. Quantum mechanics introduces extra energies that add new contributions to the familiar classical capacitance $C_g$  determined by the geometry: namely, $ C^{-1}_{tot} = C^{-1}_{g} + C^{-1}_{dos} + C^{-1}_{xc}$  where $dos$ denotes density of state, and $xc$ denotes exchange and correlations.  The first correction term is caused by the kinetic energy of the electrons. Adding electrons to a conductor requires finite kinetic energy and therefore this first contribution  $C_{dos}$   reduces the total capacitance. The second contribution results from the correlated motion of electrons, which generally leads to reduction of their total electrostatic energy. This adds a {\it negative capacitance} term $C_{xc}$  that increases the total capacitance. 
In one dimension, the capacitance plays a special role as it also determines the properties of the excitations.  Described within the Luttinger model, the fundamental excitations are collective waves of spin or charge. This electrostatic effect is captured by the compressibility of the electronic gas, or equivalently by its capacitance. Thus, the central parameter of the Luttinger liquid theory is directly related to the capacitance  by the simple relation that the Luttinger parameter $g = \sqrt{C_{tot}/C_{dos}}$." 
Again we see the deeper significance of these thermodynamic quantities. 

The above cases  exemplify how macroscopic  quantities can be judiciously used to reveal the workings of the microscopic constituents, a better understanding of which can serve as an inspiration for  the hydrodynamic and thermodynamics approaches to probing the microscopic structures of spacetime.

\subsection{Methods, Findings and Organization}

Zeta function method is used in  our calculations of the thermodynamic quantities for a thermal Bose gas (finite temperature quantum scalar field).  The main targets are $C_V, C_P$ related to the energy density fluctuations at constant volume and pressure,  and the isothermal $\kappa_T$ and adiabatic compressibility  $\kappa_S$ related to the momentum density fluctuations at finite and zero temperatures.  For completeness we also provide the energy density, entropy and the expansion coefficients $\alpha$.  Thermal fields  (with periodic imaginary time) in the following static background spacetimes are considered:  a) 2d Casimir geometry and 2d Einstein Cylinder (periodic in one spatial dimension);  b) 4d Einstein Universe $S^1 \times S^3$ ,  c)  Einstein universes with even-spatial dimensions $S^1 \times S^2$  and $S^1 \times S^4$.   

The results for the thermodynamic quantities at low and high temperatures are presented in two tables in the last section.  Please refer there for detailed  explanations.   In Sec. 2, we derive  the thermodynamic quantities in 2d (1-d space) thermal Casimir and 2d Einstein cylinder.  In Sec. 3, in 4d (3-sphere space) Einstein universe, and in Sec. 4, in 3d (2-sphere) and 5d (4-sphere) Einstein universes. In Sec. 5 we conclude with a summary and some discussions.  In this paper we focus more on the technical aspects, aiming to get a complete tally of these thermodynamic quantities. We shall continue to explore their physical significance to  better understand  how the thermodynamics of quantum fields is governed by the underlying spacetime structure and what we can say about the nature of gravity  from the  thermodynamics of quantum fields.

\newpage

\section{2d: Thermal Casimir and Einstein Cylinder}

To illustrate the points we would like to make,  on the capacities and the compressibilities of quantum fields in spaces with various topologies, we start with the simple case of an Einstein cylinder at finite temperature \cite{Boyle03,LCH16}. The topology is basically $S^1 \times S^1$.  (The first entry denotes time, the second entry denotes spatial dimensions.  Since we shall be working with thermal field in the imaginary time formulation, the first entry will always generically be $S^1$. Only in certain specified limit  will $R^1$ appear.) The length of the Euclidean time circle is characterized by $\beta=1/T$ where $T$ is the temperature. The spatial part is in a Casimir setting where the circle is characterized by the length $L$. In the low temperature limit, $L/\beta\rightarrow 0$, one has the $R^1 \times S^1$ topology with the leading contribution coming from the Casimir effect. On the other hand, in the high temperature limit, $\beta/L\rightarrow 0$, one has the thermal field in two spacetime with the topology $S^1 \times R^1$.

We shall consider a thermal minimally coupled massless scalar field in the Einstein cylinder with periodic boundary condition. To derive the Helmholtz free energy we start with the partition function, with the notation of Phillips and Hu \cite{PH00},
\begin{eqnarray}
Z=e^{W}
\end{eqnarray}
where using the proper-time zeta function method,
\begin{eqnarray}
W&=&-\frac{1}{2}{\rm Tr\ ln}\left(\frac{H}{\mu}\right)\nonumber\\
&=&\lim_{s\rightarrow 0}\frac{1}{2}\frac{d}{ds}\left[\frac{\mu^{s}}{\Gamma(s)}\int_{0}^{\infty}dt\,t^{s-1}\,{\rm Tr}(e^{-tH})\right]\nonumber\\
&=&\lim_{s\rightarrow 0}\frac{1}{2}\frac{d}{ds}\left[\frac{\mu^{s}}{\Gamma(s)}\int_{0}^{\infty}dt\,t^{s-1}\sum_{n}e^{-t\lambda_{n}}\right]\label{effact}
\end{eqnarray}
with the operator
\begin{eqnarray}
H=-\Box=-\frac{\partial^{2}}{\partial\tau^{2}}-\frac{\partial^{2}}{\partial x^{2}}
\end{eqnarray}
where $\tau$ is the Euclidean time, the eigenvalues
\begin{eqnarray}
\lambda_{n_{0},n}=k_{0}^{2}+k^{2}\ \ \ ;\ \ \ k_{0}=\frac{2\pi n_{0}}{\beta}\ \ \ ;\ \ \ k=\frac{2\pi n}{L}
\end{eqnarray}
and the eigenfunctions
\begin{eqnarray}
\phi_{n_{0},n}(x)=\left(\frac{1}{\sqrt{\beta}}e^{ik_{0}\tau}\right)\left(\frac{1}{\sqrt{L}}e^{ikx}\right).
\end{eqnarray}
With the partition function~$Z$, we proceed to calculate the free energy $F$. 
\begin{eqnarray}
Z=e^{-\beta F}\Rightarrow F&=&-\frac{W}{\beta}\nonumber\\
&=&\left(-\frac{1}{2\beta}\right)\lim_{s\rightarrow 0}\frac{d}{ds}\left[\frac{\mu^{s}}{\Gamma(s)}\int_{0}^{\infty}dt\,t^{s-1}\sum_{n_{0},n}e^{-tk_{0}^{2}}e^{-tk^{2}}\right]\nonumber\\
\label{freeF}
\end{eqnarray}
To proceed we consider the low and the high temperature expansions of this free energy. Then we can discuss the properties of various thermodynamic quantities, notably the capacities and the compressibilities that can be derived from $F$.

%\subsection{Thermal Field and Casimir Space: $S^1 \times R^1$}
\subsection{Low temperature expansion: Approaching $R^1\times S^1$}

To develop the low temperature expansion, we first rewrite the free energy $F$ using the Poisson summation formula (see, for example, \cite{Cam90}) on the sum over $n_{0}$.
\begin{eqnarray}
\sum_{n_{0}=-\infty}^{\infty}e^{-t(2\pi n_{0}/\beta)^{2}}&=&\frac{\beta}{2\sqrt{\pi t}}\sum_{n_{0}=-\infty}^{\infty}e^{-n_{0}^{2}\beta^{2}/4t},\label{Poisson}%\\
%\sum_{n=-\infty}^{\infty}e^{-t(2\pi n/L)^{2}}&=&\frac{L}{2\sqrt{\pi t}}\sum_{n=-\infty}^{\infty}e^{-n^{2}L^{2}/4t}\label{Poisson}
\end{eqnarray}
With this replacement, the free energy can then be expressed as
\begin{eqnarray}
F=\left(-\frac{1}{4\sqrt{\pi}}\right)\lim_{s\rightarrow 0}\frac{d}{ds}\left[\frac{\mu^{s}}{\Gamma(s)}\int_{0}^{\infty}dt\,t^{s-\frac{3}{2}}\sum_{n_{0},n} e^{-n_{0}^{2}\beta^{2}/4t}\,e^{-t\left(\frac{2\pi n}{L}\right)^{2}}\right]
%F=\left(-\frac{L}{8\pi}\right)\lim_{s\rightarrow 0}\frac{d}{ds}\left[\frac{\mu^{s}}{\Gamma(s)}\int_{0}^{\infty}dt\,t^{s-2}\sum_{n_{0},n}\!{\vphantom{\sum}}'\, e^{-n_{0}^{2}\beta^{2}/4t}e^{-n^{2}L^{2}/4t}\right]
\end{eqnarray}
%where the prime in the sum indicates that the $n_{0}=n=0$ term is omitted. This term corresponds to the zero temperature, flat space contribution which should be subtracted in the renormalization process.

To start with, we look at the $n_{0}=n=0$ term. This term is formally divergent so we regularize it by adding in a mass $m$.
\begin{eqnarray}
F|_{n_{0}=n=0}&=&-\frac{1}{4\sqrt{\pi}}\lim_{m\rightarrow 0, s\rightarrow 0}\frac{d}{ds}\bigg[\frac{\mu^{s}}{\Gamma(s)}
\int_{0}^{\infty}dt\,t^{s-\frac{3}{2}}\,e^{-tm^{2}}
\bigg]\nonumber\\
&=&-\frac{1}{4\sqrt{\pi}}\lim_{m\rightarrow 0}\left(-2\sqrt{\pi}\,m\right)\nonumber\\
&=&0.\label{n0neq01}
\end{eqnarray}
We shall therefore neglect this term in the following.

For the $n_{0}\neq 0$ and $n=0$ term, we have the finite temperature zero mode contribution to the free energy.
\begin{eqnarray}
F|_{n_{0}\neq 0,n=0}&=&\left(-\frac{1}{2\sqrt{\pi}}\right)\lim_{s\rightarrow 0}\frac{d}{ds}\left[\frac{\mu^{s}}{\Gamma(s)}\int_{0}^{\infty}dt\,t^{s-\frac{3}{2}}\sum_{n_{0}=1}^{\infty}\, e^{-n_{0}^{2}\beta^{2}/4t}\right]\nonumber\\
&=&-\frac{1}{2\sqrt{\pi}}\lim_{s\rightarrow 0}\frac{d}{ds}\left[\frac{\mu^{s}}{\Gamma(s)}2^{1-2s}\left(\beta^{2}\right)^{-\frac{1}{2}+s}\Gamma(\frac{1}{2}-s)\zeta(1-2s)\right]\nonumber\\
&=&\frac{1}{2\beta}\,{\rm ln}(\mu\beta^{2}),\label{neq0}
\end{eqnarray}
which is proportional to ${\rm ln}\beta/\beta$. Whether this term should be included in evaluating the free energy has been a controversial issue \cite{BMO02,Dowker03,ET02}. In \cite{BMO02}, it was argued that it should be excluded. The main reason is that the entropy derived from this term would go like ${\rm ln}\beta$ which will blow up in the zero temperature limit, being at variance with the third law of thermodynamics. On the contrary, the inclusion of the zero mode was advocated in \cite{Dowker03}, particularly to preserve the inversion relation between free energies at low and high temperatures \cite{Cardy91}. In a more mundane argument, the inclusion  of zero mode is necessary for the mode function to form a complete set, which in turns enforces causality. When the zero mode is missing, it has been shown that the lightcone structure emerges in massless relativistic field theory can be violated~\cite{SY,Edurado}.
Here, we shall include this term for completeness and from past experiences. The zero mode in the spectrum of an invariant operator governs the infrared behavior of quantum fields in curved spacetimes (see, e.g., \cite{HuOC87}). The infrared behavior of massless minimally-coupled interacting quantum field in de Sitter universe is an important problem in cosmology (see, e.g., \cite{IRdSRev} for a review where earlier references can be found). In our subsequent discussions we shall explore its consequences by investigating the properties of the corresponding thermodynamic quantities derived from it. One further remark is that this zero mode contribution existent in all compact spaces we have considered is independent of the details of the spatial geometry. Therefore, as we shall see in the following, this term will contribute to the free energy in the low temperature expansion for all spatial configurations with a zero mode in the eigenspectrum. However, in contrast to the interacting quantum field case mentioned above, for free fields considered here,  there is no zero mode contribution in the zero temperature free energy expression.  
%Note that from this term one can derive the Stefan's law in two spacetime dimensions as we shall show in the subsequent discussions. 

For $n_{0}=0$ and $n\neq 0$, 
\begin{eqnarray}
F|_{n_{0}=0,n\neq 0}&=&\left(-\frac{1}{2\sqrt{\pi}}\right)\lim_{s\rightarrow 0}\frac{d}{ds}\left[\frac{\mu^{s}}{\Gamma(s)}\int_{0}^{\infty}dt\,t^{s-\frac{3}{2}}\sum_{n=1}^{\infty}\,e^{-t\left(\frac{2\pi n}{L}\right)^{2}}\right]\nonumber\\
&=&-\frac{\pi}{6L}\label{n0eq0}
\end{eqnarray}
As $n_{0}=0$, we have the zero temperature limit here. The result actually corresponds to the quantum Casimir effect \cite{Casimir,Ford,DowCri} due to the spatial circle. This is consistent with the usual derivation of the Casimir energy. With periodic boundary condition, the allowed frequencies are just $\omega_{n}=2\pi n/L$. The vacuum energy is thus
\begin{eqnarray}
E=\sum_{n=-\infty}^{\infty}\frac{1}{2}\omega_{n} %\nonumber\\
=\left(\frac{2\pi}{L}\right)\zeta(-1) %\nonumber\\
=-\frac{\pi}{6L}\label{CasimirE}
\end{eqnarray}
which is the same as what we had above.

%For the $n_{0}\neq 0$, $n\neq 0$ terms, we shall consider the low temperature and high temperature expansions separately to obtain closed form expressions.

%Here we develope the low temperature expansion for the $n_{0}\neq 0$, $n\neq 0$ contributions to the free energy. At low temperature, $L/\beta\ll 1$. We rescale $t\rightarrow t\beta^{2}$ and expand the exponential in powers of $L/\beta$.
For $n_{0}\neq 0$, $n\neq 0$, we have
%\begin{eqnarray}
%&&F|_{n_{0}\neq 0,n\neq 0}\nonumber\\
%&=&\left(-\frac{L}{2\pi\beta^{2}}\right)\lim_{s\rightarrow 0}\frac{d}{ds}\bigg[\frac{(\mu\beta^{2})^{s}}{\Gamma(s)}\sum_{n_{0},n=1}^{\infty}
%\sum_{l=0}^{\infty}\frac{1}{l!}\left(-\frac{n^{2}L^{2}}{4\beta^{2}}\right)^{l} \int_{0}^{\infty}dt\,t^{s-2-l}e^{-n_{0}^{2}/4t}\bigg]
%\nonumber\\
%&=&\left(-\frac{L}{2\pi\beta^{2}}\right)\lim_{s\rightarrow 0}\frac{d}{ds}\bigg[\frac{(\mu\beta^{2})^{s}}{\Gamma(s)}
%\sum_{l=0}^{\infty}\frac{1}{l!}\left(-\frac{\beta^{2}}{4L^{2}}\right)^{l}\nonumber\\
%&&\hskip 100pt  2^{2+2l-2s}\Gamma(1-l-s)\zeta(-2l)\zeta(2+2l-2s)\bigg]
%\nonumber\\
%&=&\frac{\pi L}{6\beta^{2}}+\cdots\label{n0neq1l}
%\end{eqnarray}
\begin{eqnarray}
F|_{n_{0}\neq 0,n\neq 0}&=&\left(-\frac{1}{\sqrt{\pi}}\right)\lim_{s\rightarrow 0}\frac{d}{ds}\left[\frac{\mu^{s}}{\Gamma(s)}\int_{0}^{\infty}dt\,t^{s-\frac{3}{2}}\sum_{n_{0},n=1}^{\infty} e^{-n_{0}^{2}\beta^{2}/4t}\,e^{-t\left(\frac{2\pi n}{L}\right)^{2}}\right]\nonumber\\
&=&\sum_{n_{0},n=1}^{\infty}\left(-\frac{2}{n_{0}\beta}\right)e^{-2\pi n_{0}n\beta/L}\nonumber\\
&=&-\frac{2}{\beta}e^{-2\pi\beta/L}+\cdots,\label{n0neq1l}
\end{eqnarray}
where the ellipsis represents terms which are also exponentially small and non-analytic like $e^{-2\pi\beta/L}$.

Hence, combining expressions in Eqs.~(\ref{n0neq01}) to (\ref{n0neq1l}), the free energy in the low temperature expansion is given by
\begin{eqnarray}
F=-\frac{\pi}{6L}+\frac{1}{2\beta}{\rm ln}(\mu\beta^{2})+\cdots.\label{F2l}
\end{eqnarray}
Indeed, when $\beta\rightarrow\infty$ or $T\rightarrow 0$, we are left with the first term which is just the Casimir energy.

With the free energy we are able to derive various thermodynamic quantities including the heat capacities and the compressibilities \cite{MR00} in the low temperature limit. First, the entropy
\begin{eqnarray}
%S=\beta^{2}\left(\frac{\partial F}{\partial \beta}\right)_{L}\sim e^{-2\pi\beta/L}+\cdots
S=1-\frac{1}{2}{\rm ln}(\mu\beta^{2})+\cdots.\label{ZMentropy}
\end{eqnarray}
Other than the exponentially small terms, the contribution to the entropy comes solely from the zero mode since the first term corresponding to the Casimir effect in the free energy in Eq.~(\ref{F2l}) is independent of temperature. As we have mentioned above, the ${\rm ln}\beta$ term diverges as the temperature $T\rightarrow 0$, apparently violating the third law of thermodynamics in the traditional settings, under the assumptions of large spatial volume and high temperatures. This is an important issue which deserves closer examinations,  especially in the context of quantum thermodynamics in spacetimes with curvature or nontrivial topology.  We shall therefore leave this term as it is and try to explore more its consequences on the other thermodynamic quantities.

%which shows that the entropy is exponentially small in this limit. As temperature $T\rightarrow 0$ or $\beta\rightarrow\infty$, the entropy $S$ goes to zero in accordance with the third law of thermodynamics. This is so because in the low temperature limit, other than exponentially small terms, the free energy $F$ consists of only the Casimir term which is independent of temperature.

With the entropy $S$, we have the internal energy
\begin{eqnarray}
E=F+TS=-\frac{\pi}{6L}+\frac{1}{\beta}+\cdots,
\end{eqnarray}
which is independent of the scale $\mu$. The dominant term of the internal energy comes from the Casimir effect as given by Eq.~(\ref{CasimirE}). The subdominant term, which is linear in $T$, is the zero mode contribution.
The corresponding energy density would then be
\begin{eqnarray}
\rho=\frac{E}{L}=-\frac{\pi}{6L^{2}}+\frac{1}{\beta L}+\cdots.
\end{eqnarray}

From the free energy, pressure $P$ is just
\begin{eqnarray}
P=-\left(\frac{\partial F}{\partial L}\right)_{T}=-\frac{\pi}{6L^{2}}+\cdots.\label{2dpressurelow}
\end{eqnarray}
We thus have negative pressure coming from the Casimir effect. The magnitude increases as the size $L$ gets smaller. Negative pressure would shrink the spatial circle and this shrinking force would get larger as the size of circle gets smaller. This process is very similar to that of a gravitational collapse. In fact, we shall see in the subsequent discussions, the Casimir effects of spatial spheres with different dimensions would all induce negative pressure and same kind of collpases should therefore occur.

From the entropy, we obtain the heat capacity at constant volume which is given by the second temperature derivative of the Helmholtz free energy:
\begin{eqnarray}
C_{V}=-\beta\left(\frac{\partial S}{\partial\beta}\right)_{L}=1+\cdots
\end{eqnarray}
Hence, other than exponentially small terms, $C_{V}$ is basically a constant.

Furthermore, the second derivative of the free energy with respect to volume $L$ is related to the isothermal compressibility $\kappa_{T}$, 
\begin{eqnarray}
\kappa_{T}=-\frac{1}{L}\left(\frac{\partial L}{\partial P}\right)=-\frac{3L^{2}}{\pi}+\cdots.\label{2disocomlow}
\end{eqnarray}
$\kappa_{T}$ is negative due to the negative pressure of the Casimir effect. This means that when pressure is increased, or the magnitude of the pressure is decreased, the volume $L$ will increase. This is in contrast to the case of a normal gas where the volume would decrease when the pressure is increased with positive compressibility.

There is another second derivative of the free energy
\begin{eqnarray}
\frac{\partial^{2}F}{\partial T\partial L}=-\frac{\partial S}{\partial L}=-\frac{\partial P}{\partial T}
\end{eqnarray}
according to the Maxwell relations. This second derivative is related to the thermal expansion coefficient defined by
\begin{eqnarray}
\alpha=\frac{1}{L}\left(\frac{\partial L}{\partial T}\right)_{P}.
\end{eqnarray}
By the cyclic relation
\begin{eqnarray}
\left(\frac{\partial L}{\partial T}\right)_{P}&=&-\left(\frac{\partial L}{\partial P}\right)_{T}\left(\frac{\partial P}{\partial T}\right)_{L}\nonumber\\
&=&L\kappa_{T}\left(\frac{\partial P}{\partial T}\right)_{L}
\end{eqnarray}
Hence, the thermal expansion coefficient can be expressed as
\begin{eqnarray}
\alpha=\kappa_{T}\left(\frac{\partial P}{\partial T}\right)_{L}.
\end{eqnarray}
In our present consideration, with the pressure $P$ in Eq.~(\ref{2dpressurelow}) having a leading term independent of temperature, we have
\begin{eqnarray}
\left(\frac{\partial P}{\partial T}\right)_{L}=-\beta^{2}\left(\frac{\partial P}{\partial \beta}\right)_{L}\sim e^{-2\pi\beta/L}+\cdots
\end{eqnarray}
which is exponentially small. Thus, with the isothermal compressibility in Eq.~(\ref{2disocomlow}),
\begin{eqnarray}
\alpha\sim e^{-2\pi\beta/L}+\cdots
\end{eqnarray}
and it is also exponentially suppressed.

Using the thermodynamic quantities we have obtained, we can also derive the heat capacity at constant pressure $C_{P}$. By applying the cyclic and the Maxwell relations, the following relation can be established \cite{LL80}.
\begin{eqnarray}
C_{P}&=&C_{V}+\beta^{3} L\kappa_{T}\left(\frac{\partial P}{\partial \beta}\right)^{2}=1+\cdots.\label{CpCv}
\end{eqnarray}
Since $\partial P/\partial \beta$ is exponentially small, we can see that $C_{P}\sim C_{V}$ up to order $e^{-2\pi\beta/L}$.

Finally one can also obtain the adiabatic compressibility
\begin{eqnarray}
\kappa_{S}=-\frac{1}{L}\left(\frac{\partial L}{\partial P}\right)_{S}
\end{eqnarray}
Again using the Maxwell and the cyclic relations, we have the identity \cite{LL80}
\begin{eqnarray}
\frac{\kappa_{S}}{\kappa_{T}}=\frac{C_{V}}{C_{P}}\Rightarrow\kappa_{S}=\left(\frac{C_{V}}{C_{P}}\right)\kappa_{T}.\label{CRatio}
\end{eqnarray}
As we have seen that $C_{P}\sim C_{V}$ up to order $e^{-2\pi\beta/L}$. Therefore, we also have $\kappa_{S}\sim\kappa_{T}$ other than exponentially small terms. 
\begin{eqnarray}
\kappa_{S}=-\frac{1}{L}\left(\frac{\partial L}{\partial P}\right)=-\frac{3L^{2}}{\pi}+\cdots.\label{2dadicomlow}
\end{eqnarray}
Note that for a closed system, the thermodynamic processes would be adiabatic. Here, $\kappa_{S}$ being negative would induce the kind of collapses we mentioned above for a closed spatial geometry. In the subsequent sections, we shall see that this is true for the cases of different spatial geometries in the low temperature expansion.

With the considerations above, it is also possible to establish the fluctuations of various thermodynamic quantities. Consider a small part of an equilibrium system. Assume that the small part is still large enough for the thermodynamic limit to hold. According to the fluctuation theory of Landau and Lifshitz \cite{LL80,Mishin15}, one can then derive the fluctuations of the various thermodynamic quantities in this small part. Using the temperature $T$ and the volume $L$ as independent variables, we have the mean square fluctuation of the temperature
\begin{eqnarray}
\langle(\Delta T)^{2}\rangle=\frac{T^{2}}{C_{V}}=\frac{1}{\beta^{2}C_{V}},
\end{eqnarray}
while the fluctuation of the volume
\begin{eqnarray}
\langle(\Delta L)^{2}\rangle=LT|\kappa_{T}|=\frac{L|\kappa_{T}|}{\beta}.
\end{eqnarray}
As the fluctuation should be a positive quantity and in our case $\kappa_{T}$ is actually negative, we therefore define the fluctuation to be related to the absolute value of $\kappa_{T}$ instead.

Then the fluctuation of the internal energy is given by
\begin{eqnarray}
\langle(\Delta E)^{2}\rangle&=&\frac{C_{V}}{\beta^{2}}+\frac{L|\kappa_{T}|}{\beta}\left[-\beta\frac{\partial P}{\partial\beta}-P\right]^{2}\nonumber\\
&=&\frac{\pi}{12\beta L}+\frac{1}{\beta^{2}}+\cdots
\end{eqnarray}
which is proportional to $T$. Note that as $C_{V}$ is a constant and $\partial P/\partial \beta$ are exponentially small, the leading contribution comes from $L|\kappa_{T}|P^{2}/\beta$.

Again according to the Landau-Lifshitz fluctuation theory, the fluctuation of the pressure is given by
\begin{eqnarray}
\langle(\Delta P)^{2}\rangle&=&\frac{1}{\beta L|\kappa_{S}|}=\frac{\pi}{3\beta L^{3}}+\cdots
\end{eqnarray}
which is also proportional to $T$.

Moreover, the correlated fluctuation is then
\begin{eqnarray}
\langle(\Delta E)(\Delta P)\rangle=\frac{P}{\beta}=-\frac{\pi}{6\beta L^{2}}+\cdots
\end{eqnarray}
We can see that the leading behaviors of these fluctuations are all proportional to $T$. In other words, as $T\rightarrow 0$, all the fluctuations will vanish. It is therefore apparent that the fluctuations derived above are all of thermal nature. No quantum fluctuations are included in this theory of fluctuations.

%\subsection{Thermal Field in Einstein Cylinder: $S^1 \times S^1$}
\subsection{High temperature expansion: Approaching $S^1\times R^1$}

For the high temperature expansion, we take $\beta/L\ll 1$. Actually, we can also view this as a large $L$ expansion or the infinite space limit. In this case, we %would rescale $t\rightarrow tL^{2}$ and expand the exponential in powers of $\beta/L$.
implement the Poisson summation formula on the spatial sum over $n$,
\begin{eqnarray}
\sum_{n=-\infty}^{\infty}e^{-t(2\pi n/L)^{2}}&=&\frac{L}{2\sqrt{\pi t}}\sum_{n=-\infty}^{\infty}e^{-n^{2}L^{2}/4t}\label{PShigh}
\end{eqnarray}
Then the free energy becomes
\begin{eqnarray}
F=\left(-\frac{L}{4\sqrt{\pi}\beta}\right)\lim_{s\rightarrow 0}\frac{d}{ds}\left[\frac{\mu^{s}}{\Gamma(s)}\int_{0}^{\infty}dt\,t^{s-\frac{3}{2}}\sum_{n_{0},n}e^{-t\left(\frac{2\pi n_{0}}{\beta}\right)^{2}}e^{-n^{2}L^{2}/4t}\right].\nonumber\\
\end{eqnarray}

Again the $n_{0}=0$, $n=0$ term will be neglected as in the low temperature case. For $n_{0}\neq 0$, $n=0$, we have
\begin{eqnarray}
F|_{n_{0}\neq 0,n=0}&=&\left(-\frac{L}{2\sqrt{\pi}\beta}\right)\lim_{s\rightarrow 0}\frac{d}{ds}\left[\frac{\mu^{s}}{\Gamma(s)}\int_{0}^{\infty}dt\,t^{s-\frac{3}{2}}\sum_{n_{0}=1}^{\infty}e^{-t\left(\frac{2\pi n_{0}}{\beta}\right)^{2}}\right]\nonumber\\
&=&-\frac{\pi L}{6\beta^{2}}
\end{eqnarray}
which represents the flat space limit $L\rightarrow\infty$. Note that after implementing the Poisson summation formula in Eq.~(\ref{PShigh}), the $n=0$ term here does not correspond to the zero mode we discussed in the previous subsection.

For $n_{0}=0$, $n\neq 0$,
\begin{eqnarray}
F|_{n_{0}=0,n\neq 0}&=&\left(-\frac{L}{2\sqrt{\pi}\beta}\right)\lim_{s\rightarrow 0}\frac{d}{ds}\left[\frac{\mu^{s}}{\Gamma(s)}\int_{0}^{\infty}dt\,t^{s-\frac{3}{2}}\sum_{n=1}^{\infty}e^{-n^{2}L^{2}/4t}\right]\nonumber\\
&=&\frac{1}{2\beta}\,{\rm ln}\left(\mu L^{2}\right).
\end{eqnarray}
Finally, for $n_{0}\neq 0$, $n\neq 0$,
%\begin{eqnarray}
%&&F|_{n_{0}\neq 0,n\neq 0}\nonumber\\
%&=&\left(-\frac{1}{2\pi L}\right)\lim_{s\rightarrow 0}\frac{d}{ds}\bigg[\frac{(\mu L^{2})^{s}}{\Gamma(s)}\sum_{n_{0},n=1}^{\infty}
%\sum_{l=0}^{\infty}\frac{1}{l!}\left(-\frac{n_{0}^{2}\beta^{2}}{4L^{2}}\right)^{l} \int_{0}^{\infty}dt\,t^{s-2-l}e^{-n^{2}/4t}\bigg]
%\nonumber\\
%&=&\left(-\frac{1}{2\pi L}\right)\lim_{s\rightarrow 0}\frac{d}{ds}\bigg[\frac{(\mu L^{2})^{s}}{\Gamma(s)}
%\sum_{l=0}^{\infty}\frac{1}{l!}\left(-\frac{\beta^{2}}{4L^{2}}\right)^{l}\nonumber\\
%&&\hskip 100pt  2^{2+2l-2s}\Gamma(1-l-s)\zeta(-2l)\zeta(2+2l-2s)\bigg]
%\nonumber\\
%&=&\frac{\pi}{6L}+\cdots.\label{n0neq1h}
%\end{eqnarray}
\begin{eqnarray}
F|_{n_{0}\neq 0,n\neq 0}&=&\left(-\frac{L}{\sqrt{\pi}\beta}\right)\lim_{s\rightarrow 0}\frac{d}{ds}\left[\frac{\mu^{s}}{\Gamma(s)}\int_{0}^{\infty}dt\,t^{s-\frac{3}{2}}\sum_{n_{0},n=1}^{\infty}e^{-t\left(\frac{2\pi n_{0}}{\beta}\right)^{2}}e^{-n^{2}L^{2}/4t}\right]\nonumber\\
&=&\sum_{n_{0},n=1}^{\infty}\left(-\frac{2}{n\beta}\right)e^{-2\pi n_{0}n L/\beta}\nonumber\\
&=&-\frac{2}{\beta}\,e^{-2\pi L/\beta}+\cdots
\end{eqnarray}
which consist of exponentially small terms.
Hence, for high temperature, the free energy 
\begin{eqnarray}
F=-\frac{\pi L}{6\beta^{2}}+\frac{1}{2\beta}\,{\rm ln}\left(\mu L^{2}\right)\cdots\label{F2h}
\end{eqnarray}
with the ellipsis indicating terms which are exponentially small like $e^{-2\pi L/\beta}$. The leading term corresponds to the flat space $L\rightarrow\infty$ finite temperature result.

It is interesting to note that from this free energy at high temperature in Eq.~(\ref{F2h}), if we exchange $\beta\leftrightarrow L$,
\begin{eqnarray}
\beta F|_{high}\rightarrow -\frac{\pi\beta}{6L}+\frac{1}{2}{\rm ln}(\mu \beta^{2})+\cdots=\beta F|_{low},
\end{eqnarray}
including the exponentially small terms, where $F|_{low}$ is the free energy at low temperature in Eq.~(\ref{F2l}). $\beta\leftrightarrow L$ means $\beta/L\leftrightarrow L/\beta$, that is, the free energy in this case possesses an inversion symmetry between the low and the high temperatures. This symmetry is related to the Cardy formula \cite{Cardy91} and the contribution of the zero mode is crucial for this relation to hold.

As in the low temperature expansion, various thermodynamic quantities follow from the free energy in Eq.~(\ref{F2h}).
The entropy
\begin{eqnarray}
S=\frac{\pi L}{3\beta}-\frac{1}{2}{\rm ln}\left(\mu L^{2}\right)+\cdots
\end{eqnarray}
which is proportional to $T$ in the leading term.

The internal energy
\begin{eqnarray}
E=\frac{\pi L}{6\beta^{2}}+\cdots
\end{eqnarray}
which is independent of the scale $\mu$. Note that the leading term gives the flat space energy density
\begin{eqnarray}
\rho|_{L\rightarrow\infty}=\left(\frac{E}{L}\right)_{L\rightarrow\infty}=\frac{\pi}{6\beta^{2}}.
\end{eqnarray}

The pressure
\begin{eqnarray}
P=\frac{\pi}{6\beta^{2}}-\frac{1}{\beta L}+\cdots\label{EinCyP}
\end{eqnarray}
Here the pressure is positive. As $L\rightarrow\infty$, $P=\rho$ as it should be for an one-dimensional massless relativistic ideal gas.

It is interesting to note that in the low temperature limit, the dominant contribution to $P$ is the Casimir pressure which is negative. Here in the high temperature limit, the pressure is positive instead. Therefore, there must be a temperature at which the pressure vanishes. Whether the resulting configuration is stable or not depends on the sign of the compressibility. This is what we shall further explore in the following consideration.

The heat capacity at constant volume
\begin{eqnarray}
C_{V}=\frac{\pi L}{3\beta}+\cdots,
\end{eqnarray}
is proportional to $T$ in the leading behavior. 

%As the leading term in $P$ is independent of temperature, $\partial P/\partial\beta$ is exponentially small. Since $\kappa_{T}$ is proportional to the inverse of $\partial P/\partial\beta$, it is then exponentially large like $e^{2\pi L/\beta}$. Furthermore, both $\alpha$ and $C_{P}$ are proportional to $\kappa_{T}$ and they are also exponentially large.

The isothermal compressibility
\begin{eqnarray}
\kappa_{T}=-\beta L+\cdots,
\end{eqnarray}
where the ellipsis again represents exponentially small terms. Here, although the pressure is positive, the isothermal compressibility is negative. That is, at constant temperature, when the volume $L$ is decreased, the pressure also decreases. 
Remember $\kappa_{T}$ is related to $\partial P/\partial L$. Since the first term in Eq.~(\ref{EinCyP}) for $P$ is independent of $L$, the leading contribution of $\kappa_{T}$ comes from the second term. Hence, we have the peculiar situation in which the sign of $P$, that is, whether the pressure is positive or negative, is determined by the first term in the high temperature expansion, while the sign of the isothermal compressibility $\kappa_{T}$ is determined by the second term. Therefore, $\kappa_{T}$ could be positive or negative no matter what the sign of $P$ is. This is different from what we encounter in the low temperature expansion where $P$ is always negative due to the Casimir effect and the corresponding $\kappa_{T}$ is also negative from this negative pressure.

%We have just seen that the isothermal compressibility $\kappa_{T}$ is exponentially large but this is not so for the adiabatic compressibility $\kappa_{S}$, 
%\begin{eqnarray}
%\kappa_{S}=-\frac{1}{L}\left(\frac{\partial L}{\partial P}\right)_{S}.
%\end{eqnarray}
%To find $\kappa_{S}$ in our case, we note that at constant entropy $S_{0}$,
%\begin{eqnarray}
%&&S=\frac{\pi L}{3\beta}+\cdots=S_{0}\nonumber\\
%&\Rightarrow&
%L=\left(\frac{3S_{0}}{\pi}\right)\beta+\cdots
%\end{eqnarray}
%Then $L$ is basically proportional to $\beta$ up to terms which are exponentially small, and
%\begin{eqnarray}
%\left(\frac{\partial P}{\partial L}\right)_{S}&=&\left(\frac{\partial P}{\partial \beta}\right)_{S}\left(\frac{\partial\beta}{\partial L}\right)_{S}\nonumber\\
%&=&-\frac{\pi^{2}}{9S_{0}\beta^{3}}+\cdots
%\end{eqnarray}
%Hence, the adiabatic compressibility can be expressed as
%\begin{eqnarray}
%\kappa_{S}=\frac{3\beta^{2}}{\pi}+\cdots.
%\end{eqnarray}

With the isothermal compressibility $\kappa_{T}$, we can derive the thermal expansion coefficient
\begin{eqnarray}
\alpha=-\frac{\pi L}{3}+\beta+\cdots,
\end{eqnarray}
which is also negative as $\kappa_{T}$. This has the peculiar behavior that, at constant pressure, the volume $L$ will decrease with increase in temperature. 

Using the relation in Eq.~(\ref{CpCv}), we obtain the heat capacity at constant pressure
\begin{eqnarray}
C_{P}=-\frac{\pi^{2}L^{2}}{9\beta^{2}}+\frac{\pi L}{\beta}-1+\cdots,
\end{eqnarray}
which is negative even though $C_{V}$ is positive.

Finally, from the identity in Eq.~(\ref{CRatio}), we have the adiabatic compressibility
\begin{eqnarray}
\kappa_{S}=\frac{3\beta^{2}}{\pi}\left[1+\frac{9\beta}{\pi L}+\frac{72\beta^{2}}{\pi^{2}L^{2}}+O(\frac{\beta^{3}}{{L}^{3}})\right],
\end{eqnarray}
which is positive in contrast to a negative $\kappa_{T}$. If we view the quantum field in the spatial circle as a closed system, then its evolution satisfies the adiabatic condition. A positive $\kappa_{S}$ means that when the volume decreases, the pressure will increase in this adiabatic situation. Therefore, a stable state may be reached inspite of a negative $\kappa_{T}$.

Furthermore, we can also consider the fluctuations of various thermodynamic quantities. First, the fluctuation of the internal energy 
\begin{eqnarray}
\langle(\Delta E)^{2}\rangle&=&\frac{\pi^{2}L^{2}}{36\beta^{4}}+\frac{\pi L}{3\beta^{3}}+\cdots,
\end{eqnarray}
which is proportional to the $T^{4}$. 
%is proportional to the isothermal compressibility $\kappa_{T}$ which is exponentially large as we have discussed above, so is $\langle(\Delta E)^{2}\rangle$. 
Next, the pressure fluctuation is given by
\begin{eqnarray}
\langle(\Delta P)^{2}\rangle=\frac{\pi}{3\beta^{3}L}-\frac{3}{\beta^{2} L^{2}}+\frac{3}{\pi\beta L^{3}}+\cdots
\end{eqnarray}
which is proportional to $T^{3}$. Also, 
\begin{eqnarray}
\langle(\Delta E)(\Delta P)\rangle=\frac{\pi}{6\beta^{3}}-\frac{1}{\beta^{2}L}+\cdots
\end{eqnarray}
which is again proportional to $T^{3}$. As in the high temperature limit, $E\sim L/\beta^{2}$ and $P\sim 1/\beta^2$, we have the ratios $\langle(\Delta E)^{2}\rangle/E^{2}$ of the order of 1, and both $\langle(\Delta P)^{2}\rangle/P^{2}$ and $\langle(\Delta E)(\Delta P)\rangle/EP$ proportional to $\beta/L$.

\section{4d: Einstein universe  $S^1\times S^{3}$}
For thermal fields in the Einstein universe, the topology of spacetime is $S^{1}\times S^{3}$. The metric can be written as
\begin{eqnarray}
ds^{2}=d\tau^{2}+a^{2}d\bar{\Omega}_{3}^{2}
\end{eqnarray}
where $\bar{\Omega}_{3}$ is the solid angle of a three sphere, and $a$ is the ``radius'' characterizing the size of the sphere. The operator $H$ in Eq.~(\ref{effact}) is now given by
\begin{eqnarray}
H=-\frac{\partial^{2}}{\partial\tau^{2}}-\frac{1}{a^{2}}\bar{\Box},
\end{eqnarray}
where $\bar{\Box}$ is the Laplacian on $S^{3}$. The eigenvalue of $\bar{\Box}$ on $S^{3}$ is \cite{RO84}
\begin{eqnarray}
\bar{\lambda}_{n}=-n(n+2),
\end{eqnarray}
with degeneracy
\begin{eqnarray}
\bar{D}_{n}=(n+1)^{2}.
\end{eqnarray}
Hence, the eigenvalue of $H$ is
\begin{eqnarray}
\lambda=k_{0}^{2}+\left(\frac{1}{a^{2}}\right)n(n+2).
\end{eqnarray}
The corresponding eigenfunctions are
\begin{eqnarray}
\phi_{n_{0},n}(x)=\left(\frac{1}{\sqrt{\beta}}e^{ik_{0}\tau}\right)\left(\frac{1}{\sqrt{a^{3}}}Y_{3}(\Omega)\right),
\end{eqnarray}
where $Y_{3}(\Omega)$ is the hyperspherical harmonics on $S^{3}$ \cite{WA84}. Using the eigenvalues, one can establish the free energy, similar to Eq.~(\ref{freeF}), for the Einstein universe as
\begin{eqnarray}
F&=&-\frac{1}{2\beta}\lim_{s\rightarrow 0}\frac{d}{ds}\bigg[\frac{\mu^{s}}{\Gamma(s)}
\int_{0}^{\infty}dt\,t^{s-1}\sum_{n_{0}=-\infty}^{\infty}e^{-t\left(\frac{2\pi n_{0}}{\beta}\right)^{2}}\nonumber\\
&&\hskip 100pt \sum_{n=0}^{\infty}(n+1)^{2}\,e^{-\left(\frac{t}{a^{2}}\right)n(n+2)}\bigg]\label{EinF}
\end{eqnarray}
As in the analysis for the Einstein cylinder case in the last section, we shall consider the low and the high temperature expansions separately \cite{DK78,Dowker84,BKMR11}.

\subsection{Low temperature expansion}
In the low temperature expansion, it is appropriate to first rewrite the sum over $n_{0}$ using the Poisson summation formula as in Eq.~(\ref{Poisson}). The free energy is therefore expressed as
\begin{eqnarray}
F&=&-\frac{1}{4\sqrt{\pi}}\lim_{s\rightarrow 0}\frac{d}{ds}\bigg[\frac{\mu^{s}}{\Gamma(s)}
\int_{0}^{\infty}dt\,t^{s-\frac{3}{2}}\sum_{n_{0}=-\infty}^{\infty}e^{-n_{0}^{2}\beta^{2}/4t}\nonumber\\
&&\hskip 100pt \sum_{n=0}^{\infty}(n+1)^{2}\,e^{-\left(\frac{t}{a^{2}}\right)n(n+2)}\bigg].\label{FEinlow}
\end{eqnarray}

%% %%%%% This part has been used in the cylinder case 

%To start with, we look at the $n_{0}=n=0$ term. This term is formally divergent so we regularize it by adding in a mass $m$.
%\begin{eqnarray}
%F|_{n_{0}=n=0}&=&-\frac{1}{4\sqrt{\pi}}\lim_{m\rightarrow 0, s\rightarrow 0}\frac{d}{ds}\bigg[\frac{\mu^{s}}{\Gamma(s)}
%\int_{0}^{\infty}dt\,t^{s-\frac{3}{2}}\,e^{-tm^{2}}
%\bigg]\nonumber\\
%&=&\lim_{m\rightarrow 0}\left(\frac{m}{2}\right)\nonumber\\
%&=&0.
%\end{eqnarray}
%We shall therefore neglect this term in the following.

%%%%%%%%

To start with, we look at the $n_{0}=n=0$ term. This term is exactly the same as in Eq.~(\ref{n0neq01}) which is regularized to zero. We shall therefore neglect it in the following.
For $n_{0}= 0$ and $n\neq 0$ in Eq.~(\ref{FEinlow}), we evaluate the $t$-integral to give
\begin{eqnarray}
&&F|_{n_{0}=0, n\neq 0}\nonumber\\
&=&\lim_{s\rightarrow 0}\frac{d}{ds}\left[\left(-\frac{1}{4\sqrt{\pi}}\right)\left(\frac{\mu^{s}\Gamma(s-\frac{1}{2})}{\Gamma(s)}\right)\sum_{n=1}^{\infty}(n+1)^{2}\left(\frac{a^{2}}{n(n+2)}\right)^{s-\frac{1}{2}}\right]\nonumber\\
\end{eqnarray}
Here we concentrate on the sum
\begin{eqnarray}
\sum_{n=1}^{\infty}(n+1)^{2}\left(\frac{1}{n(n+2)}\right)^{s-\frac{1}{2}}=\sum_{n=0}^{\infty}(n+2)^{2}\left(\frac{1}{(n+3)(n+1)}\right)^{s-\frac{1}{2}}
\end{eqnarray}
Using the Plana summation formula
\begin{eqnarray}
\sum_{n=0}^{\infty}f(n)=\int_{0}^{\infty}dx\,f(x)+\frac{1}{2}f(0)+i\int_{0}^{\infty}dy\left(\frac{f(iy)-f(-iy)}{e^{2\pi y}-1}\right),\label{Plana}
\end{eqnarray}
it is possible to evalute this sum as an analytic function of $s$ \cite{OHS83,Eli95}.
\begin{eqnarray}
\sum_{n=1}^{\infty}(n+1)^{2}\left(\frac{1}{n(n+2)}\right)^{s-\frac{1}{2}}=\left(-\frac{1}{16}\right)\frac{1}{s}-0.411461+\cdots\label{EinLowSum}
\end{eqnarray}
With these considerations, we have the $n_{0}=0$ part of the free energy,
\begin{eqnarray}
F|_{n_{0}=0}=-\frac{0.224909}{a}-\frac{1}{32a}\,{\rm ln}\left(\mu a^{2}\right)
\end{eqnarray}
which represents the contribution from the Casimir effect. This result is exactly the same as the effective potential obtained in \cite{OHS83}. The dependence on $\mu$ in the free energy occurs in all Einstein universes with spatial odd spheres. The corresponding zeta function at $s=0$ is nonzero and hence contributes to a ln$\mu$ term.

For $n_{0}\neq 0$ and $n=0$, we evaluate the $t$-integral to give
\begin{eqnarray}
F|_{n_{0}\neq 0,n=0}&=&\left(-\frac{1}{2\sqrt{\pi}}\right)\lim_{s\rightarrow 0}\frac{d}{ds}\left[\left(\frac{\mu^{s}}{2^{2s-1}\Gamma(s)}\right)\Gamma(\frac{1}{2}-s)\zeta(1-2s)\right]\nonumber\\
&=&\frac{1}{2\beta}\,{\rm ln}(\mu\beta^{2}).
\end{eqnarray}
This is the contribution from the zero mode as we have discussed in the last section. This same expression will be present in the low temperature expansions of the free energy in all the cases we consider in this paper.

For $n_{0}\neq 0$ and $n\neq 0$, we again evaluate the $t$-integral first to obtain
\begin{eqnarray}
F|_{n_{0}\neq 0, n\neq 0}&=&\sum_{n_{0},n=1}^{\infty}\left(-\frac{(n+1)^{2}}{n_{0}\beta}\right)e^{\sqrt{n(n+2)}\,n_{0}\beta/a}\nonumber\\
&=&-\frac{4}{\beta}e^{-\sqrt{3}\beta/a}+\cdots
\end{eqnarray}
The $n_{0}\neq 0$ and $n\neq 0$ contribution to the free energy in the low energy expansion is exponentially small, and combining the various parts we have
\begin{eqnarray}
F=-\frac{0.224909}{a}-\frac{1}{32a}{\rm ln}(\mu a^{2})+\frac{1}{2\beta}{\rm ln}(\mu\beta^{2})+\cdots\label{EinFLow}
\end{eqnarray}
where the ellipsis represents exponentially small terms. 

With the above Helmholtz free energy in the low temperature expansion, we derive the various thermodynamic quantities. For the entropy 
\begin{eqnarray}
S=1-\frac{1}{2}\,{\rm ln}(\mu\beta^{2})+\cdots
\end{eqnarray}
which is the entropy of the zero mode with the same expression as in Eq.~(\ref{ZMentropy}).

%with a dependence on the renormalization scale $\mu$. This dependence on $\mu$ indicates that the thermodynamics of a quantum field in the Einstein universe involves divergent quantities and a renormalization procedure is needed to define the various physically measureable quantities. Here, the definition of the entropy $S$ and also of the other thermodynamic functions, as we shall see in the following discussions, would require the determination of $\mu$ through some physical scale in the problem. This is also the case for the internal energy

For the internal energy
\begin{eqnarray}
E=-\frac{0.224909}{a}-\frac{1}{32a}{\rm ln}(\mu a^{2})+\frac{1}{\beta}+\cdots,\label{EinELow}
\end{eqnarray}
and the energy density, with the volume of the three sphere $V=4\pi^{2}a^{3}$,
\begin{eqnarray}
\rho=\frac{E}{V}=-\frac{0.00570}{a^4}-\frac{1}{128\pi^{2}a^{4}}{\rm ln}(\mu a^{2})+\frac{1}{4\pi^{2}a^{3}\beta}+\cdots.\label{EinRhoLow}
\end{eqnarray}
The leading behavior in the low temperature expansion comes from the first two terms. They constitute the Casimir contribution which is $\mu$ dependent. Moreover, the pressure
\begin{eqnarray}
P=-\frac{0.00274258}{a^{4}}-\frac{1}{192\pi^{2}a^{4}}{\rm ln}(\mu a^{2})+\cdots\label{EinPLow}
\end{eqnarray}
which is negative due to the Casimir terms and is also dependent on $\mu$.

From the entropy $S$, we derive the heat capacity at constant volume
\begin{eqnarray}
C_{V}=1+\cdots
\end{eqnarray}
which is basically a constant with the temperature terms exponentially suppressed. From the pressure $P$, we obtain the isothermal compressibility
\begin{eqnarray}
\kappa_{T}=-a^{4}\left[0.00330496+\frac{1}{144\pi^{2}}{\rm ln}(\mu a^{2})\right]^{-1}+\cdots\label{EinKTLow}
\end{eqnarray}
which is negative and dependent on $\mu$. Again, the temperature-dependent terms are exponentially suppressed. Since the temperature-dependent terms in $P$ are exponentially small, the thermal expansion coefficient $\alpha$ which depends on $\partial P/\partial T$ is therefore exponentially small too. For the same reason, the heat capacity at constant pressure $C_{P}\sim C_{V}$ up to exponentially small terms. The same applies to the two compressibilities $\kappa_{S}\sim\kappa_{T}$.

With $C_{V}$ and $\kappa_{T}$, we can derive the fluctuation of the internal energy
\begin{eqnarray}
\langle(\Delta E)^{2}\rangle=\frac{2\pi^{2}}{\beta a}\left[\frac{(0.00274258+0.000527714\,{\rm ln}(\mu a^{2}))^{2}}{0.00330496+0.000703619\,{\rm ln}(\mu a^{2})}\right]+\frac{1}{\beta^{2}}+\cdots
\end{eqnarray}
which is proportional $T$. Note that since $\kappa_{T}$ is negative and the fluctuation should be a positive quantity, we have taken the absolute value of $\kappa_{T}$ in the formula for $\langle(\Delta E)^{2}\rangle$.
The fluctuation of $P$ is related to the adiabatic compressibility $\kappa_{S}$. Again we take its absolute value in the formula.
\begin{eqnarray}
\langle(\Delta P)^{2}\rangle=\frac{1}{\beta a^{7}}\left[0.000167431+0.0000356458\, {\rm ln}(\mu a^{2})\right]+\cdots
\end{eqnarray}
which is also proportional to $T$ and dependent on the renormalization scale. Moreover, the correlation between the fluctuations of $E$ and $P$,
\begin{eqnarray}
\langle(\Delta E)(\Delta P)\rangle=-\frac{1}{\beta a^{4}}\left[0.00274258+0.000527714\, {\rm ln}(\mu a^{2})\right]+\cdots
\end{eqnarray}

It is interesting to see that here all the thermodynamic quantities and their fluctuations depend on the renormalization scale $\mu$. This dependence indicates that the thermodynamics of a quantum field in the Einstein universe involves divergent quantities and a renormalization procedure is needed to define the various physically measureable quantities. As we shall see in the following section,  this is not the case with even spatial $S^{2}$ and $S^{4}$  dimensional spheres. Actually, the dependence on $\mu$ comes from the fact that we have an odd-dimensional sphere ($S^{3}$) here. We would expect the same kind of dependence to occur for all odd spheres or (when adding in the time dimension)  even-dimensional spacetimes \cite{DN92}.

\subsection{High temperature expansion}
In this subsection, we concentrate on the high temperature expansion for the Einstein universe with $\beta/a\ll 1$. In this case, we consider the Helmholtz free energy in Eq.~(\ref{EinF}). Here, we would neglect the $n_{0}=n=0$ term. This term is independent of the temperature $T$ as well as the size $a$ of the spatial three sphere. Although it is formally divergent, it should be subtracted in the renormalization procedure. 

For the $n_{0}=0$, $n\neq 0$ contribution to the free energy,
\begin{eqnarray}
F_{n_{0}=0,n\neq 0}&=&-\frac{1}{2\beta}\lim_{s\rightarrow 0}\frac{d}{ds}\bigg[\frac{\mu^{s}}{\Gamma(s)}
\int_{0}^{\infty}dt\,t^{s-1}\sum_{n=1}^{\infty}(n+1)^{2}\,e^{-\left(\frac{t}{a^{2}}\right)n(n+2)}\bigg]\nonumber\\
&=&-\frac{1}{2\beta}\lim_{s\rightarrow 0}\frac{d}{ds}\left\{(\mu a^{2})^{s}\sum_{n=1}^{\infty}(n+1)^{2}[n(n+2)]^{-s}\right\}.
\end{eqnarray}
We apply the Plana summation formula, as in Eq.~(\ref{Plana}), to the sum
\begin{eqnarray}
\sum_{n=1}^{\infty}(n+1)^{2}[n(n+2)]^{-s}&=&\sum_{n=0}^{\infty}(n+2)^{2}[(n+1)(n+3)]^{-s}\nonumber\\
&=&-1-1.20563\,s+\cdots
\end{eqnarray}
This shows that the sum as an analytic function of $s$ is well behaved near $s=0$. The $n_{0}=0$, $n\neq 0$ contribution to the free energy can therefore be written as
\begin{eqnarray}
F_{n_{0}=0,n\neq 0}&=&\frac{0.60282}{\beta}+\frac{1}{2\beta}{\rm ln}(\mu a^{2}).\label{Fn00}
\end{eqnarray}

Next, we consider the $n_{0}\neq 0$ part of the free energy
\begin{eqnarray}
F|_{n_{0}\neq 0}&=&-\frac{1}{\beta}\lim_{s\rightarrow 0}\frac{d}{ds}\bigg[\frac{\mu^{s}}{\Gamma(s)}
\int_{0}^{\infty}dt\,t^{s-1}\sum_{n_{0}=1}^{\infty}e^{-t\left(\frac{2\pi n_{0}}{\beta}\right)^{2}}\nonumber\\
&&\hskip 100pt \sum_{n=0}^{\infty}(n+1)^{2}\,e^{-\left(\frac{t}{a^{2}}\right)n(n+2)}\bigg]\label{n0neq0}
\end{eqnarray}
We concentrate on the sum over $n$. In the high temperature limit, we are interested in the asymptotic behavior of this sum \cite{BKM88} when $a$ is large or as a power series in $t/a$. To develop such a series, we again use the Plana summation formula in Eq.~(\ref{Plana}).
\begin{eqnarray}
&&\sum_{n=0}^{\infty}(n+1)^{2}\,e^{-\left(\frac{t}{a^{2}}\right)n(n+2)}\nonumber\\
&=&\left(\frac{a}{\sqrt{t}}\right)^{3}\left[\frac{\sqrt{\pi}}{4}+\frac{\sqrt{\pi}}{4}\left(\frac{t}{a}\right)+\frac{\sqrt{\pi}}{8}\left(\frac{t}{a}\right)^{2}+\cdots\right].
\end{eqnarray}
With this asymptotic expansion, one can evaluate Eq.~(\ref{n0neq0}) as a power series in $\beta/a$.
\begin{eqnarray}
F|_{n_{0}\neq 0}=-\frac{\pi^{4}a^{3}}{45\beta^{4}}-\frac{\pi^{2}a}{12\beta^{2}}-\frac{1}{32a}\left[2\gamma+{\rm ln}\left(\frac{\mu\beta^{2}}{16\pi^{2}}\right)\right]+\cdots\label{Fn0n0}
\end{eqnarray}
where $\gamma$ is the Euler constant.

With Eqs.~(\ref{Fn00}) and (\ref{Fn0n0}), the free energy in the high temperature expansion is then given by
\begin{eqnarray}
F&=&-\frac{\pi^{4}a^{3}}{45\beta^{4}}-\frac{\pi^{2}a}{12\beta^{2}}+\frac{0.60282}{\beta}\nonumber\\
&&\ \ \ \ \ +\frac{1}{2\beta}{\rm ln}(\mu a^{2})-\frac{1}{32a}\left[2\gamma+{\rm ln}\left(\frac{\mu\beta^{2}}{16\pi^{2}}\right)\right]+\cdots%\nonumber\\
\end{eqnarray}
From this free energy, we again derive the various thermodynamic quantities.

For the entropy,
\begin{eqnarray}
S=\frac{4\pi^{4}a^{3}}{45\beta^{3}}+\frac{\pi^{2}a}{6\beta}-\frac{1}{2}{\rm ln}(\mu a^{2})-0.60282-\frac{\beta}{16a}+\cdots
\end{eqnarray}
which is proportional to $T^{3}$ in the leading behavior. Note that it is dependent on the renormalization scale $\mu$ but only in the subleading term. For the internal energy,
\begin{eqnarray}
E=\frac{\pi^{4}a^{3}}{15\beta^{4}}+\frac{\pi^{2}a}{12\beta^{2}}-\frac{1}{32a}\left[2+2\gamma+{\rm ln}\left(\frac{\mu\beta^{2}}{16\pi^{2}}\right)\right]+\cdots.
\end{eqnarray}
With the volume of the three sphere as $V=2\pi^{2}a^{3}$, the first term of the internal energy $E$ is $\pi^{2}V/30\beta^{4}$ which is again the Stefan's law in flat spatial three dimensions. Indeed, in the infinite space limit, $a\rightarrow \infty$, the energy density
\begin{eqnarray}
\rho|_{a\rightarrow\infty}=\frac{E|_{a\rightarrow\infty}}{V}=\frac{\pi^{2}}{30\beta^{4}}.
\end{eqnarray}
As for the pressure,
\begin{eqnarray}
P=\frac{\pi^{2}}{90\beta^{4}}+\frac{1}{72\beta^{2}a^{2}}-\frac{1}{6\pi^{2}\beta a^{3}}-\frac{1}{192\pi^{2}a^{4}}\left[2\gamma+{\rm ln}\left(\frac{\mu\beta^{2}}{16\pi^{2}}\right)\right]+\cdots.\nonumber\\
\label{EinP}
\end{eqnarray}
The first term gives the pressure in the infinite space limit,
\begin{eqnarray}
P|_{a\rightarrow\infty}=\frac{\pi^{2}}{90\beta^{4}}=\frac{\rho|_{a\rightarrow\infty}}{3}
\end{eqnarray}
which is the equation of state for a massless relativistic ideal gas in flat three spatial dimensions. 

For the heat capacity at constant volume
\begin{eqnarray}
C_{V}=\frac{4\pi^{4}a^{3}}{15\beta^{3}}+\frac{\pi^{2}a}{6\beta}+\frac{\beta}{16a}+\cdots
\end{eqnarray}
which is proportional to $T^{3}$ in the leading term and is independent of $\mu$. Furthermore, the isothermal compressibility $\kappa_{T}$ is
%, the thermal expansion coefficient $\alpha$, the heat capacity at constant pressure $C_{P}$, and the adiabatic compressibility $\kappa_{S}$ are, respectively,
\begin{eqnarray}
\kappa_{T}&=&108\beta^{2}a^{2}+\left(\frac{1944}{\pi^{2}}\right)\beta^{3}a+\frac{81\beta^{4}}{\pi^{2}}\left[\frac{432}{\pi^{2}}+2\gamma+{\rm ln}\left(\frac{\mu\beta^{2}}{16\pi^{2}}\right)\right]+\cdots.\nonumber\\
\end{eqnarray}

%% This part has been used in the high temperature expansion of the Einstein cylinder
%Remember $\kappa_{T}$ is related to $\partial P/\partial V$ or $\partial P/\partial a$. Since the first term in Eq.~(\ref{EinP}) for $P$ is independent of $a$, the leading contribution of $\kappa_{T}$ comes from the second term. Hence, we have the peculiar situation in which the sign of $P$, that is, whether the pressure is positive or negative, is determined by the first term in the high temperature expansion, while the sign of the isothermal compressibility $\kappa_{T}$ is determined by the second term. Therefore, $\kappa_{T}$ could be positive or negative no matter what the sign of $P$ is. This is different from what we encounter in the low temperature expansion where $P$ is always negative due to the Casimir effect and the corresponding $\kappa_{T}$ is also negative from this negative pressure.

From the isothermal compressibility, one can obtain the thermal expansion coefficient $\alpha$, the heat capacity at constant pressure $C_{P}$, and the adiabatic compressibility $\kappa_{S}$. They are, respectively,
\begin{eqnarray}
\alpha&=&\frac{24\pi^{2}a^{2}}{5\beta}+\frac{432\,a}{5}+\frac{18\beta}{5}\left[\frac{5}{6}+\frac{432}{\pi^{2}}+2\gamma+{\rm ln}\left(\frac{\mu\beta^{2}}{16\pi^{2}}\right)\right]+\cdots.\nonumber\\
\\
%\end{eqnarray}
%The heat capacity at constant pressure is
%\begin{eqnarray}
C_{P}&=&\frac{32\pi^{6}a^{5}}{75\beta^{5}}+\frac{192\pi^{4}a^{4}}{25\beta^{4}}+\frac{8\pi^{4}a^{3}}{25\beta^{3}}\left[\frac{5}{2}+\frac{432}{\pi^{2}}+2\gamma+{\rm ln}\left(\frac{\mu\beta^{2}}{16\pi^{2}}\right)\right]+\cdots.\nonumber\\
\\
%\end{eqnarray}
%The adiabatic compressibility is
%\begin{eqnarray}
\kappa_{S}&=&\frac{135\beta^{4}}{2\pi^{2}}-\frac{675\beta^{6}}{8\pi^{4}a^{2}}+\frac{10125\beta^{7}}{8\pi^{6}a^{3}}+\cdots.
\end{eqnarray}
The relationship between the heat capacities is $C_{P}=C_{V}+\beta^{3}V\kappa_{T}(\partial P/\partial\beta)^{2}$. In the present high temperature expansion, the second term which is of the order of $(a/\beta)^{5}$ dominates over the first term which is only of the order $(a/\beta)^{3}$. Hence, we have $C_{P}\gg C_{V}$. For the same reason, since $\kappa_{S}=(C_{V}/C_{P})\kappa_{T}$, we have $\kappa_{T}\gg \kappa_{S}$.

Next, we give the fluctuations for various thermodynamic quantities in the high temperature expansion. For the internal energy fluctuation,
\begin{eqnarray}
&&\langle(\Delta E)^{2}\rangle\nonumber\\
&=&\frac{6\pi^{6}a^{5}}{25\beta^{7}}+\frac{108\pi^{4}a^{4}}{25\beta^{6}}+\frac{9\pi^{4}}{50\beta^{5}}\left[\frac{70}{27}+\frac{432}{\pi^{2}}+2\gamma+{\rm ln}\left(\frac{\mu\beta^{2}}{16\pi^{2}}\right)\right]+\cdots.\nonumber\\
\end{eqnarray}
For the pressure fluctuation,
\begin{eqnarray}
\langle(\Delta P)^{2}\rangle=\frac{1}{135\beta^{5}a^{3}}+\frac{1}{108\pi^{2}\beta^{3}a^{5}}-\frac{5}{36\pi^{4}\beta^{2}a^{6}}+\cdots.
\end{eqnarray}
Lastly, the correlated fluctuation of $E$ and $P$ is
\begin{eqnarray}
\langle(\Delta E)(\Delta P)\rangle=\frac{\pi^{2}}{90\beta^{5}}+\frac{1}{72\beta^{3}a^{2}}-\frac{1}{6\pi^{2}\beta^{2}a^{3}}+\cdots.
\end{eqnarray}
We see that in this high temperature expansion, $\langle(\Delta E)^{2}\rangle$ is proportional to $T^7$, while both $\langle(\Delta P)^{2}\rangle$ and $\langle(\Delta E)(\Delta P)\rangle$ are proportional to $T^{5}$.

%\section{: $S^1\times S^n$}
\section{Even spatial dimensions: $S^1\times S^{2}$ and $S^1\times S^{4}$}
In the last section we have considered the Einstein universe with a spatial three sphere. The consideration there can be extended to Einstein universes with the general topology of $S^{1}\times S^{d-1}$ with a spatial $(d-1)$-sphere. The metric is given by
\begin{eqnarray}
ds^{2}=d\tau^{2}+a^{2}d\bar{\Omega}_{d-1}^{2}
\end{eqnarray}
where $\bar{\Omega}_{d-1}$ is the solid angle of the $(d-1)$-sphere. Here the eigenvalue of the Laplacian $\bar{\Box}$ on $S^{d-1}$ is given by \cite{RO84}
\begin{eqnarray}
\bar{\lambda}_{n}=-n(n+d-2),
\end{eqnarray}
and the degeneracy is
\begin{eqnarray}
\bar{D}_{n}=\frac{(2n+d-2)(n+d-3)!}{n!(d-2)!}
\end{eqnarray}
Using $\bar{\lambda}_{n}$ and $\bar{D}_{n}$, one can write the Helmholtz free energy in this generalized Einstein universe as
\begin{eqnarray}
F=-\frac{1}{2\beta}\lim_{s\rightarrow 0}\frac{d}{ds}\bigg[\frac{\mu^{s}}{\Gamma(s)}
\int_{0}^{\infty}dt\,t^{s-1}\sum_{n_{0}=-\infty}^{\infty}e^{-t\left(\frac{2\pi n_{0}}{\beta}\right)^{2}}
\sum_{n=0}^{\infty}\bar{D}_{n}e^{-\left(\frac{t}{a^{2}}\right)\bar{\lambda}_{n}}\bigg].\label{Fndim}
\end{eqnarray}
With analyses similar to those in the last section, one can derive the corresponding free energy $F$ in both the low temperature and the high temperature expansions for any value of the dimension $d$. Then, one can derive the various thermodynamic quantities from the free energy. In the following subsections, we shall work out the $d=3$ and $d=5$ cases explicitly. Together with the case $d=4$ in the last section, we can have a better understanding in how the various thermodynamic quantities depend on the spacetime dimension \cite{Dowker84,DN92}.

\subsection{Low temperature expansion}
In the low temperature expansion with $a/\beta\ll 1$, we   first rewrite the sum over $n_{0}$ in Eq.~(\ref{Fndim}) using the Poisson summation formula, as in Eq.~(\ref{FEinlow}) for the 4d Einstein universe case. Then, following similar considerations there, we obtain the free energies for the $d=3$ and $d=5$ Einstein universe cases.
\begin{eqnarray}
F|_{d=3}&=&-\frac{0.132548}{a}+\frac{1}{2\beta}{\rm ln}(\mu\beta^{2})+\cdots,\nonumber\\
F|_{d=5}&=&-\frac{0.215872}{a}+\frac{1}{2\beta}{\rm ln}(\mu\beta^{2})+\cdots,
\end{eqnarray}
where the ellipsis represents terms which are exponentially small. Comparing with the free energy of $d=4$ Einstein universe in Eq.~(\ref{EinFLow}), we see that in the free energies above, the term ${\rm ln}(\mu a^{2})/a$ is missing. This can be traced back to the sum in Eq.~(\ref{EinLowSum}), where the sum, as an analytic function of $s$, has a pole at $s=0$. This is characteristic for this kind of sums on unit odd dimensional spheres. On the other hand, for even spheres like $S^{2}$ and $S^{4}$, the corresponding sums would be power series in $s$ without pole singularities at $s=0$. For this reason, the free energies above for the $d=3$ and $d=5$ Einstein universes do not contain terms like ${\rm ln}(\mu a^{2})/a$. Subsequently, as we shall see below, the compressibilities $\kappa_{T}$ and $\kappa_{S}$, and the thermal expansion coefficient $\alpha$, together with the various fluctuations, will not depend on the renormalization scale $\mu$.

Using these free energies, we derive the entropies
\begin{eqnarray}
S|_{d=3}&=&1-\frac{1}{2}{\rm ln}(\mu\beta^{2})+\cdots,\nonumber\\
S|_{d=5}&=&1-\frac{1}{2}{\rm ln}(\mu\beta^{2})+\cdots.
\end{eqnarray}
Both have a dependence on the renormalization scale $\mu$. From the entropies, we have the heat capacities at constant volume,
\begin{eqnarray}
C_{V}|_{d=3}&=&1+\cdots,\nonumber\\
C_{V}|_{d=5}&=&1+\cdots.
\end{eqnarray}
which are constant except for exponentially small terms. The corresponding quantities of the Einstein universe in the last section have the same form as the ones above. Therefore, we would expect that this is true for all the Einstein universes with general $(d-1)$ dimensional spatial spheres.

The internal energies
\begin{eqnarray}
E|_{d=3}&=&-\frac{0.132548}{a}+\frac{1}{\beta}+\cdots,\nonumber\\
E|_{d=5}&=&-\frac{0.215872}{a}+\frac{1}{\beta}+\cdots,
\end{eqnarray}
and the corresponding energy densities
\begin{eqnarray}
\rho|_{d=3}&=&-\frac{0.0105478}{a^3}+\frac{1}{4\pi \beta a^{2}}+\cdots,\\
\rho|_{d=5}&=&-\frac{0.00820215}{a^5}+\frac{3}{8\pi^{2}\beta a^{5}}+\cdots.
\end{eqnarray}
The first term in both expressions represents the Casimir energy. Other than the exponentially small terms, there is also one term proportional to $T$. We expect Einstein universes with odd spacetimes dimensions to have the same behaviors for $E$ and $\rho$ as above. Moreover, for Einstein universes with even spacetime dimensions, there should be an extra term proportional to ${\rm ln}(\mu a^{2})/a$ as shown in Eqs.~(\ref{EinELow}) and (\ref{EinRhoLow}) in the last section.

For the pressures in these cases,
\begin{eqnarray}
P|_{d=3}&=&-\frac{0.00527392}{a^3}+\cdots,\nonumber\\
P|_{d=5}&=&-\frac{0.00205054}{a^{5}}+\cdots.
\end{eqnarray}
The Casimir pressures in both cases are negative. This is true for all the cases we have considered in the previous sections in the low temperature expansion. From the expressions for the pressures, one can derive the isothermal compressibilities
\begin{eqnarray}
\kappa_{T}|_{d=3}&=&-126.408\, a^3+\cdots,\nonumber\\
\kappa_{T}|_{d=5}&=&-390.141\,a^{5}+\cdots.
\end{eqnarray}
Both isothermal compressibilities are negative due to the negative pressure. This is also true for all the isothermal compressibilities in all the cases we have considered. However, unlike $\kappa_{T}$ in the 4d Einstein universe but similar to the Einstein cylinder case, the isothermal compressibilities here are not dependent on $\mu$. Since the pressures $P$ are dominated by the Casimir effect which is independent of temperature, the thermal expansion coefficients are exponentially small in these case. Also, for the same reason, we have $C_{P}\sim C_{V}$ and $\kappa_{S}\sim\kappa_{T}$ up to exponentially small terms.

Finally, we  layout the expressions for the various fluctuations. For the internal energy,
\begin{eqnarray}
\langle(\Delta E)^{2}\rangle|_{d=3}&=&0.0441827\left(\frac{1}{\beta a}\right)+\frac{1}{\beta^{2}}+\cdots,\nonumber\\
\langle(\Delta E)^{2}\rangle|_{d=5}&=&0.0431744\left(\frac{1}{\beta a}\right)+\frac{1}{\beta^{2}}+\cdots.
\end{eqnarray}
For the pressure,
\begin{eqnarray}
\langle(\Delta P)^{2}\rangle|_{d=3}&=&0.000629528\left(\frac{1}{\beta a^{5}}\right)+\cdots,\nonumber\\
\langle(\Delta P)^{2}\rangle|_{d=5}&=&0.0000973889\left(\frac{1}{\beta a^{9}}\right)+\cdots,
\end{eqnarray}
The fluctuations are proportional to $T$. This is true for all the cases we have considered. In addition, the correlated fluctuations of $E$ and $P$,
\begin{eqnarray}
\langle(\Delta E)(\Delta P)\rangle|_{d=3}&=&-0.00527392\left(\frac{1}{\beta a^{3}}\right)+\cdots,\nonumber\\
\langle(\Delta E)(\Delta P)\rangle|_{d=5}&=&-0.00205054\left(\frac{1}{\beta a^{5}}\right)+\cdots,
\end{eqnarray}
which are also proportional to $T$. 

Since the energies and the pressures in the low temperature limit are all dominated by the Casimir effect, their leading behaviors are independent of $T$. Therefore, the ratios $\langle(\Delta E)^{2}\rangle/E^{2}$, $\langle(\Delta P)^{2}\rangle/P^{2}$, and $\langle(\Delta E)(\Delta P)\rangle/EP$ go like $a/\beta$ or $aT$ in the low temperature limit. Fluctuations of all Einstein universes in the low temperature expansion have similar behavior, except that for even spacetimes the fluctuations will have dependences on the renormalization scale $\mu$, while here for odd spacetimes they will not.

\subsection{High temperature expansion}
In this subsection we develop the high temperature expansion with $\beta/a\ll 1$. For the Helmholtz free energies in Eq.~(\ref{Fndim}), we again use the same procedure as in the 4d Einstein universe case in Sec.~3 to obtain
\begin{eqnarray}
F|_{d=3}&=&-\frac{2\zeta(3)a^{2}}{\beta^{3}}+\left(\frac{1}{\beta}\right)\left[0.580842+\frac{1}{3}{\rm ln}(\mu a^{2})+\frac{1}{6}{\rm ln}(\mu\beta^{2})\right]\nonumber\\
&&\hskip 30pt -\frac{\beta}{360a^{2}}-\frac{\beta^{3}}{113400a^{4}}+\cdots,\nonumber\\
F|_{d=5}&=&-\frac{2\zeta(5)a^{4}}{\beta^{5}}-\frac{2\zeta(3)a^{2}}{3\beta^{3}}\nonumber\\
&&\hskip 30pt +\left(\frac{1}{\beta}\right)\left[0.276064+0.338889\,{\rm ln}(\mu a^{2})+\frac{29}{180}{\rm ln}(\mu\beta^{2})\right]\nonumber\\
&&\hskip 30pt -\frac{37\beta}{4536a^{2}}+\cdots.
\end{eqnarray}

The corresponding entropies are
\begin{eqnarray}
S|_{d=3}&=&\frac{6\zeta(3)a^{2}}{\beta^{2}}-\left[0.247509+\frac{1}{3}{\rm ln}(\mu a^{2})+\frac{1}{6}{\rm ln}(\mu\beta^{2})\right]\nonumber\\
&&\hskip 30pt -\frac{\beta^{2}}{360 a^{2}}-\frac{\beta^{4}}{37800 a^{4}}+\cdots,\nonumber\\
S|_{d=5}&=&\frac{10\zeta(5)a^{4}}{\beta^{4}}+\frac{2\zeta(3)a^{2}}{\beta^{2}}\nonumber\\
&&
%\hskip 30pt 
+\left[0.0461582-0.338889\,{\rm ln}(\mu a^{2})-\frac{29}{180}{\rm ln}(\mu\beta^{2})\right]%\nonumber\\
%&&\hskip 30pt 
-\frac{37\beta^{2}}{4536a^{2}}+\cdots.\nonumber\\
\end{eqnarray}
The internal energies are
\begin{eqnarray}
E|_{d=3}&=&\frac{4\zeta(3)a^{2}}{\beta^{3}}+\frac{0.333333}{\beta}-\frac{\beta}{180a^{2}}-\frac{\beta^{3}}{28350a^{4}}+\cdots,\nonumber\\
E|_{d=5}&=&\frac{8\zeta(5)a^{4}}{\beta^{5}}+\frac{4\zeta(3)a^{2}}{3\beta^{3}}+\frac{0.322222}{\beta}-\frac{37\beta}{2268 a^{2}}+\cdots.
\end{eqnarray}
The first terms again give the flat space limit of the internal energy. Hence, as $a\rightarrow\infty$, we have the energy density
\begin{eqnarray}
\rho|_{d=3,a\rightarrow\infty}&=&\frac{E|_{d=3,a\rightarrow\infty}}{4\pi a^{2}}=\frac{\zeta(3)}{\pi\beta^{3}},\nonumber\\
\rho|_{d=5,a\rightarrow\infty}&=&\frac{E|_{d=5,a\rightarrow\infty}}{\left(\frac{8}{3}\pi^{2}a^{4}\right)}=\frac{3\zeta(5)}{\pi^{2}\beta^{5}}.\label{rho35}
\end{eqnarray}
For the pressures
\begin{eqnarray}
P|_{d=3}&=&\frac{\zeta(3)}{2\pi\beta^{3}}-\frac{1}{12\pi\beta a^{2}}-\frac{\beta}{1440\pi a^{4}}-\frac{\beta^{3}}{226800\pi a^{6}}+\cdots,\nonumber\\
P|_{d=5}&=&\frac{3\zeta(5)}{4\pi^{2}\beta^{5}}+\frac{\zeta(3)}{8\pi^{2}\beta^{3}a^{2}}-\frac{0.00643812}{\beta a^{4}}-\frac{37\beta}{24192\pi^{2}a^{6}}+\cdots.\nonumber\\
\label{EinPHigh}
\end{eqnarray}
The first terms give the flat space limit. As $a\rightarrow\infty$, we see that $P$ will be proportional to the energy density $\rho$ in Eq.~(\ref{rho35}).
\begin{eqnarray}
P|_{d=3,a\rightarrow\infty}=\frac{\zeta(3)}{2\pi\beta^{3}}=\frac{1}{2}\rho|_{d=3,a\rightarrow\infty},\nonumber\\
P|_{d=5,a\rightarrow\infty}=\frac{3\zeta(5)}{4\pi^{2}\beta^{5}}=\frac{1}{4}\rho|_{d=5,a\rightarrow\infty}.
\end{eqnarray}
This is consistent with the equation of state of a massless relativistic ideal gas, $P=\rho/n$, in a $n$-dimensional flat space. From the entropies above, we can also work out the heat capacities at constant volume.
\begin{eqnarray}
C_{V}|_{d=3}&=&\frac{12\zeta(3)a^{2}}{\beta^{2}}+\frac{1}{3}+\frac{\beta^{2}}{180a^{2}}+\frac{\beta^{4}}{9450a^{4}}+\cdots,\nonumber\\
C_{V}|_{d=5}&=&\frac{40\zeta(5)a^{4}}{\beta^{4}}+\frac{4\zeta(3)a^{2}}{\beta^{2}}+\frac{29}{90}+\frac{37\beta^{2}}{2268 a^{2}}+\cdots.
\end{eqnarray}

From the pressures, we can derive the isothermal compressibilities,
\begin{eqnarray}
\kappa_{T}|_{d=3}&=&-12\pi\beta a^{2}+\frac{\pi\beta^{3}}{5}-\frac{\pi\beta^{5}}{700a^{2}}+\cdots,\nonumber\\
\kappa_{T}|_{d=5}&=&\frac{16\pi^{2}\beta^{3}a^{2}}{\zeta(3)}+111.109\beta^{5}+\frac{97.9843\beta^{7}}{a^{2}}+\cdots.
\end{eqnarray}
This result is a little bit surprising as we see that $\kappa_{T}|_{d=3}$ is negative in the high temperature limit, while the isothermal compressibilities for $d=4$ and $5$ are both positive in the same limit. We have discussed this point briefly in the last section on the isothermal compressiblity of the $d=4$ Einstein universe in the high temperature expansion. Here, if we look at the expressions for the pressure $P$ above in Eq.~(\ref{EinPHigh}), we see that the leading terms of $P$ are independent of $a$. Since $\kappa_{T}$ is proportional to the inverse of $\partial P/\partial a$, the sign of $\kappa_{T}$ would depend on the sign of the second terms in Eq.~(\ref{EinPHigh}). Interestingly, for $d=3$, the second term of $P$ is negative, while for $d=5$, this term is positive. In fact, although we have not detailed here, one can show that this term is negative only for $d=3$, and it is positive for all other dimensions $d\geq 4$. Hence, we have the peculiar result that $\kappa_{T}$ is negative only for the $d=3$ Einstein universe, and positive for all other Einstein universes.

Since the thermal expansion coefficient $\alpha$ is proportional to $\kappa_{T}$, we have a negative $\alpha$ for $d=3$ and positive $\alpha$ for  dimensions $d\geq 4$. Indeed, we have 
\begin{eqnarray}
\alpha|_{d=3}&=&-\frac{18\zeta(3)a^{2}}{\beta}+1.36062\beta-\frac{0.0275758\beta^{3}}{a^{2}}+\cdots,\nonumber\\
\alpha|_{d=5}&=&\frac{60\zeta(5)a^{2}}{\zeta(3)\beta}+49.7752\beta+\frac{42.8332\beta^{3}}{a^{2}}+\cdots.
\end{eqnarray}
This is also true for the heat capacities at constant pressure,
\begin{eqnarray}
C_{P}|_{d=3}&=&-\frac{108\zeta(3)^{2}a^{4}}{\beta^{4}}+\frac{31.4503a^{2}}{\beta^{2}}-0.379195+\cdots,\nonumber\\
C_{P}|_{d=5}&=&\frac{600\zeta(5)^{2}a^{6}}{\zeta(3)\beta^{6}}+\frac{619.826a^{4}}{\beta^{4}}+\frac{500.021 a^{2}}{\beta^{2}}+\cdots.
\end{eqnarray}
Note that the relationship between $C_{P}$ and $C_{V}$ is $C_{P}=C_{V}+\beta^{2}V\kappa_{T}(\partial P/\partial \beta)^{2}$. Here, in the high temperature expansion, the second term, which is proportional to $\kappa$, dominates over the first one. Hence, the sign of $C_{P}$ is determined by $\kappa_{T}$, and we have $C_{P}$ negative for $d=3$ and positive for all other dimensions $d\geq 4$.

It is interesting to see that, for the adiabatic compressibilities,
\begin{eqnarray}
\kappa_{S}|_{d=3}&=&\frac{4\pi\beta^{3}}{3\zeta(3)}+\frac{0.724734\beta^{5}}{a^{2}}+\frac{0.138007\beta^{7}}{a^{4}}+\cdots,\nonumber\\
\kappa_{S}|_{d=5}&=&\frac{16\pi^{2}\beta^{5}}{15\zeta(5)}-\frac{1.96158\beta^{7}}{a^{2}}+\frac{1.45329\beta^{9}}{a^{4}}+\cdots.
\end{eqnarray}
they are both positive. This is due to the relation $\kappa_{S}=(C_{V}/C_{P})\kappa_{T}$. For $d=3$, both $C_{P}$ and $\kappa_{T}$ are negative but the ratio is positive. Therefore, $\kappa_{S}$ is positive in both cases.

Lastly, we also list the fluctuations for $E$ and $P$ for completeness. For the energy fluctuations,
\begin{eqnarray}
\langle(\Delta E)^{2}\rangle|_{d=3}&=&\frac{69.3572a^{4}}{\beta^{6}}+\frac{13.2687a^{2}}{\beta^{4}}+\frac{0.501864}{\beta^{2}}+\cdots,\nonumber\\
\langle(\Delta E)^{2}\rangle|_{d=5}&=&\frac{343.481a^{6}}{\beta^{8}}+\frac{398.348a^{4}}{\beta^{6}}+\frac{320.334a^{2}}{\beta^{4}}+\cdots.
\end{eqnarray}
For the pressure fluctuations,
\begin{eqnarray}
\langle(\Delta P)^{2}\rangle|_{d=3}&=&\frac{0.0228363}{\beta^{4}a^{2}}-\frac{0.00474943}{\beta^{2}a^{4}}+\frac{0.0000833667}{a^{6}}+\cdots,\nonumber\\
\langle(\Delta P)^{2}\rangle|_{d=5}&=&\frac{0.00374241}{\beta^{6}a^{4}}+\frac{0.000723064}{\beta^{4}a^{6}}-\frac{0.000396002}{\beta^{2}a^{8}}+\cdots.\nonumber\\
\end{eqnarray}
For the correlated fluctuations of $E$ and $P$,
\begin{eqnarray}
\langle(\Delta E)(\Delta P)\rangle|_{d=3}&=&\frac{0.191313}{\beta^{4}}-\frac{0.0265258}{\beta^{2}a^{2}}-\frac{0.000221049}{a^{4}}+\cdots,\nonumber\\
\langle(\Delta E)(\Delta P)\rangle|_{d=5}&=&\frac{0.0787971}{\beta^{6}}+\frac{0.0152242}{\beta^{4}a^{2}}-\frac{0.00643812}{\beta^{2}a^{4}}+\cdots.\nonumber\\
\end{eqnarray}
From these results and also the ones in the last section for the $d=4$ Einstein universe, we have in the high temperature expansion, $\langle(\Delta E)^{2}\rangle$ proportional to $T^{d+3}$, and both $\langle(\Delta P)^{2}\rangle$ and $\langle(\Delta E)(\Delta P)\rangle$ proportional to $T^{d+1}$. We also have the rations $\langle(\Delta E)^{2}\rangle/E^{2}$ proportional to $(\beta/a)^{d-3}$, and both $\langle(\Delta P)^{2}\rangle/P^{2}$ and $\langle(\Delta E)(\Delta P)\rangle/EP$ proportional to $(\beta/a)^{d-1}$.

\section{Conclusions and discussions}

%The purpose of our present investigation is to knit a thermodynamic picture of these quantum field fluctuation phenomena.

In this paper,   towards the loftier goal of exploring the relationship between quantum fields, spacetimes and thermodynamics, we  take on a more manageable task,  of finding  a thermodynamic  description of the fluctuations of the stress energy of quantum fields in several generic static spacetimes. We have derived expressions for the  heat capacity at constant volume and pressure, as well as the isothermal and adiabatic compressibility of a thermal scalar quantum field in a 2d (1 time and 1 space) Casimir space, a 2d Einstein cylinder, and 3d, 4d, 5d (spatial $S^2, S^3, S^4$) Einstein universes.  We now analyze  these results considering these factors:   comparison between cases with positive and negative heat capacity and compressibility,  effects of curvature, topology and dimensionality.  Our next paper \cite{CHH2} will deal with the same issues in dynamical spacetimes relevant to cosmology.

\begin{table}
\caption{\label{table1}Low temperature leading behaviors of various thermodynamic quantities.}
\centering
\setlength{\tabcolsep}{10pt}
\renewcommand{\arraystretch}{1.5}
%\begin{ruledtabular}
\begin{tabular}%{\textwidth}
 {ccccc}
\hline\hline
 & $d=2$ & $3$ & $4$ & $5$\\ 
\hline 
$\rho$ & $-\pi/6L$ & $-0.0105478/a^3$ & Eq.~(\ref{EinRhoLow}) &$-0.00820215/a^{5}$ \\
$S$ & $1-\frac{1}{2}{\rm ln}(\mu\beta^{2})$ & $1-\frac{1}{2}{\rm ln}(\mu\beta^{2})$ & $1-\frac{1}{2}{\rm ln}(\mu\beta^{2})$ & $1-\frac{1}{2}{\rm ln}(\mu\beta^{2})$ \\ 
$C_{V}$ & $1$ & 1 & $1$ & 1 \\ 
$C_{P}$ &  $1$ & 1 & $1$ & 1\\ 
$P$ & $-\pi/6L^{2}$ & $-0.0053/a^3$ & Eq.~(\ref{EinPLow}) & $-0.0021/a^5$ \\ 
$\kappa_{T}$ &  $-3L^2/\pi$ & $-126 a^{3}$ & Eq.~(\ref{EinKTLow}) & $-390 a^{5}$ \\
$\kappa_{S}$ & $-3L^2/\pi$ & $-126 a^{3}$ & Eq.~(\ref{EinKTLow}) & $-390 a^{5}$ \\
$\alpha$ & $\sim e^{-2\pi\beta/L}$ & $\sim e^{-\sqrt{2}\beta/a}$ & $\sim e^{-\sqrt{3}\beta/a}$ & $\sim e^{-2\beta/a}$ \\
\hline\hline
\end{tabular}
%\end{ruledtabular}
\end{table}

Our results are summarized in Tables \ref{table1} and \ref{table2}. In Table \ref{table1}, the leading behaviors of various thermodynamic quantities of a minimally coupled massless scalar field in the low temperature expansion are tabulated. We can see that the energy density $\rho$, the pressure $P$, the compressibilities $\kappa_{T}$ and $\kappa_{S}$ are all negative. This is because  the leading behaviors are mostly dominated by the Casimir effect.   We therefore have the universal feature that the compressibilities in the low temperature limit, in the Einstein cylinder and all the Einstein universes studied here, are negative. In fact, the magnitude of the negative pressure is inversely proportional to the size of the spatial geometry, meaning that it would compress the spatial geometry, and while the circle or the sphere gets smaller, the shrinking pressure would get larger. The process is similar to an accelerated gravitational collapse. This is actually the mechanism behind the so-called spontaneous compacification of the extra dimensions in the Kaluza-Klein scenario \cite{AC83,CW84}. 

Since the Casimir effect is  solely controlled by the spatial geometries, the leading behavior of the free energy is independent of temperature $T$ or $\beta$. The temperature-dependent terms are subdominate. They are related to the entropy $S$ and the heat capacities $C_{V}$ and $C_{P}$ as derivatives of the free energy with respect to the temperature. 
In all cases in the low temperature expansion, the subleading contribution to the free energy comes from the zero mode with $F_{ZM}=\frac{1}{2\beta}\,{\rm ln}(\mu\beta^{2})$. From this we obtain the leading behaviors of $S$, $C_{V}$, and $C_{P}$ as follows:
\begin{eqnarray}
&&S_{ZM}=1-\frac{1}{2}\,{\rm ln}(\mu\beta^{2})+\cdots\nonumber\\
 &&(C_{V})_{ZM}=1+\cdots\ \ \ \ \ ;\ \ \ \ \ (C_{P})_{ZM}=1+\cdots
\end{eqnarray}
As we have mentioned before, $S_{ZM}$ apparently violates the third law of thermodynamics. If the zero mode is ignored in the calculation of the free energy, the temperature-dependent part of the free energy would come from exponentially small terms. Then, $S$, $C_{V}$, and $C_{P}$ would all be exponentially small.
%For $d=2$, the subleading terms are $\sim e^{-2\pi\beta/L}$, that is, they are exponentially small as shown in the table. For $d\geq 4$ Einstein universes, the subleading terms give an entropy of the form $1-\frac{1}{2}{\rm ln}(\mu \beta^{2})$ and basically constant $C_{V}$ and $C_{P}$, while the coefficient of thermal expansion $\alpha$ is exponentially small like $e^{-\beta/a}$.

Another interesting feature in Table \ref{table1} is that for the $d=4$ Einstein universe, the various thermodynamic quantities, except $\alpha$, are explicitly dependent on the renormalization scale $\mu$. In fact, as we have discussed earlier, this feature is also true for all Einstein universes with odd dimensional spatial spheres. This can be traced back to the sum like Eq.~(\ref{EinLowSum}),
\begin{eqnarray}
f(s)=\sum_{n=1}^{\infty}\bar{D}_{n}\left(\frac{1}{\bar{\lambda}_{n}}\right)^{s-\frac{1}{2}},
\end{eqnarray}
where $\bar{\lambda}_{n}$ and $\bar{D}_{n}$ are respectively the eigenvalue and the degeneracy of the Laplacian on a sphere. For odd spheres, as in Eq.~(\ref{EinLowSum}), the function $f(s)$ has a pole singularity at $s=0$, and this induces the renormalization scale dependence of the thermodynamic quantities. For even spheres, the function $f(s)$ is analytic at $s=0$ without the pole term, and hence there is no $\mu$ dependence for the thermodynamic quantities \cite{DN92,Eli95}.

\begin{table}
\caption{\label{table2}High temperature leading behaviors of various thermodynamic quantities.}
\centering
\setlength{\tabcolsep}{7pt}
\renewcommand{\arraystretch}{1.5}
%\begin{ruledtabular}
\begin{tabular}%{\textwidth}
 {ccccc}
\hline\hline
 & $d=2$ & $3$ & $4$ & $5$\\ 
\hline 
$\rho$ & $\pi/6\beta^{2}$ & $\zeta(3)/\pi\beta^{3}$ & $\pi^{2}/30\beta^{4}$ &$3\zeta(5)/\pi^{2}\beta^{5}$ \\
$S$ & $\pi L/3\beta$ & $6\zeta(3)a^{2}/\beta^{2}$ & $4\pi^{4}a^{3}/45\beta^{3}$ & $10\zeta(5)a^{4}/\beta^{4}$ \\ 
$C_{V}$ & $\pi L/3\beta$ & $12\zeta(3)a^{2}/\beta^{2}$ & $4\pi^{4}a^{3}/15\beta^{3}$ & $40\zeta(5)a^{4}/\beta^{4}$ \\ 
$C_{P}$ &  $-\pi^{2}L^{2}/9\beta^{2}$ & $-108(\zeta(3))^{2}a^{4}/\beta^{3}$ & $32\pi^{6}a^{5}/75\beta^{5}$ & $600(\zeta(5))^{2}a^{6}/\zeta(3)\beta^{5}$ \\ 
$P$ & $\pi/6\beta^{2}$ & $\zeta(3)/2\pi\beta^{3}$ & $\pi^{2}/90\beta^{4}$ &$3\zeta(5)/4\pi^{2}\beta^{5}$ \\ 
$\kappa_{T}$ &  $-\beta L$ & $-12\pi\beta a^{2}$ & $108\beta^{2}a^{2}$ & $16\pi^{2}\beta^{3}a^{2}/\zeta(3)$ \\
$\kappa_{S}$ & $3\beta^{2}/\pi$ & $4\pi\beta^{3}/3\zeta(3)$ & $135\beta^{4}/2\pi^{2}$ & $16\pi^{2}\beta^{5}/15\zeta(5)$ \\
$\alpha$ & $-\pi L/3$ & $-18\zeta(3)a^{2}/\beta$ & $24\pi^{2}a^{2}/5\beta$ & $60\zeta(5)a^{2}/\zeta(3)\beta$ \\
\hline\hline
\end{tabular}
%\end{ruledtabular}
\end{table}

In Table \ref{table2}, the leading behaviors of the thermodynamic quantities in the high temperature expansion are tabulated. One can see that the leading behaviors coincide with those of a massless relativistic ideal thermal gas in $d$-dimensional flat spacetimes, with the equation of state $P=\rho/(d-1)$. More specifically, the leading behaviors of $P$ and $\rho$ are both of the order of $T^{d}$, independent of the size of the spatial sphere. %For this reason, the isothermal compressibility $\kappa_{T}$ which is proportional to the inverse of $\partial P/\partial a$ would be related to the subleading term of $P$. For $d=2$, the subleading terms are exponentially small like $e^{-2\pi L/\beta}$. Hence, $\partial P/\partial a$ is also exponentially small but $\kappa_{T}$ would then be exponentially large like $e^{2\pi L/\beta}$ as indicated in the table. Since the thermal expansion coefficient $\alpha$ is proportional to $\kappa_{T}$, it is exponentially large as well. For $C_{P}=C_{V}+\beta^{2}V\kappa_{T}(\partial P/\partial\beta)^{2}$, the second term which contains $\kappa_{T}$ dominates over $C_{V}$ to give $C_{P}$ the exponentially large behavior.

For $d=2$ and $3$, we have the peculiar results that $\kappa_{T}$, $C_{P}$, and $\alpha$ are negative although the pressure $P$ itself is positive. This is related to the subleading terms of $P$ as shown in Eqs.~(\ref{EinCyP}) and (\ref{EinPHigh}). We can see that the leading terms are positive but the subleading terms are negative. Therefore, with $\kappa_{T}$ being related to the inverse of $\partial P/\partial a$, it is also negative. Due to this negative sign of $\kappa_{T}$, the thermal expansion coefficient $\alpha$ and the heat capacity at constant pressure $C_{P}$ would both be negative. Note that since $\kappa_{S}=(C_{V}/C_{P})\kappa_{T}$ and the signs of $\kappa_{T}$ and $C_{P}$ cancel to give a positive adiabatic compressibility $\kappa_{S}$. On the other hand, for $d\geq 4$, the subleading term of $P$, as exemplified in Eq.~(\ref{EinP}) for $d=4$ and Eq.~(\ref{EinPHigh}) for $d=5$, is again positive so we have all the thermodynamic quantities positive for all dimensions $d\geq 4$. One may wonder if this peculiar behavior of a thermal quantum field in $d=2$ Einstein cylinder and $d=3$ Einstein universe would be related to the peculiarity of  gravity  in $1+1$ and $2+1$ spacetime dimensions \cite{Brown88,Carlip98}.

In our next paper \cite{CHH2} we shall consider dynamical quantum fields and use the nonequilibrium free energy density discovered in \cite{HHNEqFE} to explore the quantum capacity and vacuum compressibility of the Universe.   \\

\noindent{\bf Acknowledgments}   HTC is supported in part by the Ministry of Science and Technology, Taiwan, ROC, under the Grants MOST109-2112-M-032-007 and MOST110-2112-M-032-009.  J.-T. Hsiang is supported by the Ministry of Science and Technology of Taiwan, R.O.C. under Grant No. MOST 110-2811-M-008-522.

\end{document}